\documentclass[aps,showpacs,twocolumn,
superscriptaddress]{revtex4}
\usepackage{graphicx}
\usepackage{dcolumn}
\usepackage{bm}
\usepackage{color}
\usepackage[normalem]{ulem} 
\usepackage[dvipsnames]{xcolor} 
\usepackage{hyperref}
\usepackage{orcidlink}
\hypersetup{
  colorlinks=true,        
  linkcolor=blue,         
  citecolor=cyan,         
}
\usepackage{mathrsfs}
\usepackage[utf8]{inputenc}
\usepackage{mathtools}
\usepackage{doi}
\usepackage{amsmath}
\usepackage{amssymb}

\usepackage{scalerel}
\usepackage{tikz}
\usetikzlibrary{svg.path}

\definecolor{orcidlogocol}{HTML}{A6CE39}
\tikzset{
  orcidlogo/.pic={
    \fill[orcidlogocol] svg{M256,128c0,70.7-57.3,128-128,128C57.3,256,0,198.7,0,128C0,57.3,57.3,0,128,0C198.7,0,256,57.3,256,128z};
    \fill[white] svg{M86.3,186.2H70.9V79.1h15.4v48.4V186.2z}
                 svg{M108.9,79.1h41.6c39.6,0,57,28.3,57,53.6c0,27.5-21.5,53.6-56.8,53.6h-41.8V79.1z M124.3,172.4h24.5c34.9,0,42.9-26.5,42.9-39.7c0-21.5-13.7-39.7-43.7-39.7h-23.7V172.4z}
                 svg{M88.7,56.8c0,5.5-4.5,10.1-10.1,10.1c-5.6,0-10.1-4.6-10.1-10.1c0-5.6,4.5-10.1,10.1-10.1C84.2,46.7,88.7,51.3,88.7,56.8z};
  }
}

\newcommand\orcidicon[1]{\href{https://orcid.org/#1}{\mbox{\scalerel*{
\begin{tikzpicture}[yscale=-1,transform shape]
\pic{orcidlogo};
\end{tikzpicture}
}{|}}}}

\begin{document}

\title{Horizon-Brightened Acceleration Radiation and the Deflection Angle Near a Degenerate Photon Sphere of Schwarzschild-like Quantum-Corrected  Black Hole}

\author{Yunqiao Xu}
\email{25B911033@stu.hit.edu.cn}
\affiliation{School of Physics, Harbin Institute of Technology, Harbin 150001, People’s Republic of China}

\author{Uktamjon Uktamov\orcidlink{0009-0003-0423-2474}} \email[Corresponding Author:]{uktam.uktamov11@gmail.com}
\affiliation{School of Physics, Harbin Institute of Technology, Harbin 150001, People’s Republic of China}
\affiliation{University of Tashkent for Applied Sciences, Str. Gavhar 1, Tashkent 100149, Uzbekistan}
\affiliation{Institute for Advanced Studies, New Uzbekistan University, Movarounnahr str. 1, Tashkent 100000, Uzbekistan}

\author{Bobomurat Ahmedov\orcidlink{0000-0002-1232-610X}} \email{ahmedov@astrin.uz}
\affiliation{School of Physics, Harbin Institute of Technology, Harbin 150001, People’s Republic of China}

\affiliation{Institute of Theoretical Physics, National University of Uzbekistan, Tashkent 100174, Uzbekistan}
\author{Chengxun Yuan} \email{yuancx@hit.edu.cn}
\affiliation{School of Physics, Harbin Institute of Technology, Harbin 150001, People’s Republic of China}

\date{\today}

\begin{abstract}
We investigate horizon-brightened acceleration radiation (HBAR) and a strong-deflection expansion for the deflection angle of light rays scattered in the vicinity of a degenerate photon sphere, within the context of a quantum-corrected black hole spacetime. We characterize the horizon structure and thermodynamics, and we extract the divergent part of the deflection-angle integral from the near-marginal-orbit contribution using a nonsingular prescription at marginality, obtaining a unique leading power-law term. In terms of the closest-approach radius, the strong-deflection leading coefficient factorizes into a universal branch constant and a local factor involving the third derivative of the effective potential at the degenerate photon sphere. On the quantum side, we develop the near-horizon reduction relevant to HBAR, demonstrating that the dominant sector governing the detector response exhibits conformal behavior and yields a thermal excitation spectrum characterized by the horizon temperature. We adopt a Lindblad master-equation framework for the radiation field, establish the existence of a thermal steady state, and obtain an HBAR entropy–energy relation that satisfies a Clausius-type first-law structure. Also, we derive a Wien-type displacement law for the HBAR spectrum, connecting the peak wavelength to horizon thermodynamics and thereby providing an additional observable probe of quantum gravity via near-horizon radiation.

\end{abstract}


\maketitle

\section{Introduction}

Over the past decade, black hole astrophysics has advanced dramatically, shifting these objects from theoretical predictions to active observational arenas for probing strong-field gravity. At the core of this framework, general relativity portrays gravity geometrically: spacetime curvature dictates the behavior of matter and light. In contrast, the energy and momentum of matter simultaneously shape the gravitational field through the field equations (\cite{Einstein:1916vd}).
 This progress has been achieved by two impressive achievements: the Event Horizon Telescope's direct imaging of the supermassive black holes M87* and Sgr A* (\cite{EventHorizonTelescope:2019dse,EventHorizonTelescope:2019ths,EventHorizonTelescope:2022wkp}), which provided the first visual confirmation of black hole shadows, and the LIGO/Virgo detection of gravitational waves from binary black hole mergers (\cite{LIGOScientific:2016aoc}), which revealed the dynamical behavior of compact binaries. Within this context, black holes stand at the heart of modern physics—they represent the most extreme classical predictions of general relativity while simultaneously serving as natural laboratories where the intricate interplay among spacetime geometry, quantum field theory, and thermodynamic principles becomes inevitable (\cite{Bardeen:1973gs}).
 
Despite the remarkable  triumphs of classical general relativity, fundamental theoretical challenges still open—most notably the longstanding conflict between classical gravitational collapse and the unavoidable emergence of curvature singularities (\cite{Penrose:1964wq,Hawking:1970zqf}), which paves a way to the need for a more complete physical description. The singularity theorems show that this kind of breakdown is inevitable under broad conditions, which means classical general relativity cannot fully describe nature at extremely high energies. To fix this, we need to incorporate quantum gravity effects into the theory.

Asymptotic safety (\cite{Reuter:1996cp,Niedermaier:2006ns,Reuter_Saueressig_2019}) is one of the most promising ideas among the many attempts to formulate quantum gravity. This approach holds that gravity becomes renormalizable without relying on perturbation theory, because a nontrivial fixed point at high energies determines how the theory behaves in that regime. A recent study (\cite{Alencar:2026hss,10.1088/1674-1137/ae66d2}) developed a concrete renormalization group improved black hole solution resembling the Schwarzschild geometry. This new solution has a distance-dependent Newton constant $G(r)$: it equals the usual $G_0$ at large distances but changes at short scales. This removes the singularity, creating a de Sitter-like core, while the spacetime looks like Schwarzschild far away.

The HBAR framework looks at a stream of tiny quantum systems—treated as two-level atoms—as they fall toward a black hole and exchange energy with the fields in the region close to the horizon (\cite{Scully:2017utk,Camblong:2020pme}). Studies of various regular BH modifications and alternative geometries have shown that changes near the horizon can impact both quantum processes like Hawking radiation and astrophysical observables such as shadows, orbital motion, QPOs, particle collisions, and accretion signatures. This strengthens the case for the combined quantum and phenomenological perspective we take here (\cite{UktamjonUktamov:2026dep,Ovgun:2025ehi}). The main idea is that near the horizon, the motion of freely falling objects and the extreme gravitational redshift together make the field interaction look very much like Unruh radiation from an accelerating observer (\cite{PhysRevD.14.870,Davies:1974th}). In this view, the horizon environment turns normally virtual excitation and emission events into real ones, producing thermal radiation along with a flow of entropy (\cite{Scully:2017utk}). This approach is helpful for two reasons: it explains horizon thermodynamics through real quantum events, and it can be applied to other black hole types—not just Schwarzschild—to see if the thermodynamic description remains valid.

Generally, in spherical symmetry the circular orbits of photons form what is known as a photon surface - meaning that once a light ray is launched tangent to this surface, it stays tangent forever (\cite{Claudel:2000yi}).  However, the authors of (\cite{Igata:2026ivq}) distinguish between two types of photon surfaces: they term radially unstable circular photon orbits a photon sphere, while radially stable ones are called an antiphoton sphere (\cite{Cvetic:2016bxi}). Only the unstable configuration generally produces a divergent deflection angle; stable circular photon orbits, in contrast, yield a finite deflection angle (\cite{Kudo:2022ewn}). At specific critical values of the spacetime parameters, a photon sphere and an antiphoton sphere may merge into a degenerate photon sphere (\cite{Hod:2017zpi,PhysRevLett.119.251102}), in which the circular photon orbits are marginally unstable. In the present work, we have examined the influence of quantum gravity corrections on the strong-deflection expansion of the light-ray deflection angle in the vicinity of a degenerate photon sphere, adopting a quantum-corrected black hole as the underlying spacetime background.

This paper is organized as follows: In Sec.\ref{Sec.I} we investigate the line element of the quantum-corrected Schwarzschild BH by analyzing event horizon radius $r_h$, Hawking temperature  $T_H$ of the quantum-corrected BH. The trajectory of the atoms falling into quantum corrected BH is investigated in Sec.\ref{Sec.II}. Section \ref{Sec.III} is devoted to applying the HBAR framework to atoms in free fall toward these quantum-corrected Schwarzschild-type black holes, with the aim of deriving the resulting radiation features in a form suitable for entropy accounting. In Sec. \ref{Sec.IV}, we construct the HBAR entropy flux and examine its relationship to the horizon area law for the spacetimes considered here. Section \ref{Sec.V} is further devoted to an exploration of the HBAR entropy, the area law, and Wien's displacement law. Section \ref{Sec.VI} is devoted to formulating the deflection angle for scattering null geodesics within the geometry of a quantum-corrected Schwarzschild-type black hole. In Sec.\ref{Sec.VII}, we extract the power-law behavior by performing a near-critical expansion of the deflection-angle integral, and we determine both the leading coefficients and the finite remainder. Sec.\ref{Sec.VIII}  contains our concluding summary and an accompanying discussion of the key results.

\section{Quantum-corrected Schwarzschild-like black hole}\label{Sec.I}
We investigate a static and spherically symmetric quantum-corrected black hole of mass $M$, with the line element given by (\cite{10.1088/1674-1137/ae66d2}):
\begin{equation}\label{eq.metric}
ds^2=-f(r)dt^2+f(r)^{-1}dr^2+r^2\left(d\theta^2 +\sin^2{\theta} d\phi^2\right),
\end{equation}
where
\begin{equation}\label{eq.metric fuction}
f(r)=1-\frac{4M r^2}{\xi^2(\gamma M+r)+\sqrt{\xi^4(\gamma M+r)^2+4r^6}}.
\end{equation}

The metric function contains two parameters: $\xi$ specifying the cutoff scale, and $\gamma$ serving as the interpolation parameter. In the limit $\xi \to 0$, the quantum corrections vanish, and the spacetime reduces to the standard Schwarzschild BH.

The event horizon radius $r_h$ is given by the equation $f(r)=0$,
\begin{equation}\label{eq.horizen}
M=\frac{r_h\xi^2\pm\sqrt{4r_h^6-2r_h^4\gamma \xi^2+r_h^2\xi^4}}{2(2r_h^2-\gamma \xi^2)}.
\end{equation}

The horizon radius is implicitly determined by Eq. (\ref{eq.horizen}); conversely, it is also convenient to express the mass $M$ as a function of $r_h$, a form that will be extensively used in our near-horizon analysis.

For the metric given in (\ref{eq.metric}), the equation $f(r)=0$ typically possesses two positive real roots (\cite{UktamjonUktamov:2025qts}), corresponding to the inner horizon (Cauchy horizon) and the outer horizon (event horizon), respectively. In black hole thermodynamics and Hawking radiation, the outer horizon is of fundamental physical quantity. Moreover, the surface gravity and the Hawking temperature are determined by the behavior of the metric function at the outer horizon. 
 
To characterize the metric function in the near-horizon region, we compute its first and second derivatives:
\begin{eqnarray}\label{eq. derivative1}
f^\prime(r)&=&\frac{4M r(\mathcal{B}_1 r-2\mathcal{A}_1)}{\mathcal{A}_1^2}\,,
\\ \label{eq. derivative2}
f^{\prime \prime}(r)&=&\frac{4M\mathcal{C}_1 r^2-8M\mathcal{A}_1}{\mathcal{A}_1^2}-\frac{8M\mathcal{B}_1 r(\mathcal{B}_1 r-2\mathcal{A}_1)}{\mathcal{A}_1^3}.
\end{eqnarray}
By evaluating the parameter on the horizon $r=r_h$ and incorporating Eq. (\ref{eq.horizen}), we obtain the following compact forms:
\begin{eqnarray}\label{eq. derivatives rh}
f^\prime_h&\equiv& f^\prime(r_h)=\frac{2r_h\mathcal{C}_2\left[r_h(\xi^2+\mathcal{A}_2)-2\mathcal{B}_2\right]}{(2r_h^2-\gamma \xi^2)\mathcal{B}_2^2}\,,
\end{eqnarray}
\begin{eqnarray}\label{eq. derivatives rh2}
\begin{split}
f^{\prime \prime}_h\equiv f^{\prime \prime}(r_h)=&-\frac{4r_h\mathcal{C}_2(\xi^2+\mathcal{A}_2)\left[r_h(\xi^2+\mathcal{A}_2)-2\mathcal{B}_2\right]}{(2r_h^2-\gamma \xi^2)\mathcal{B}_2^3}
\\&+\frac{2r_h^2\mathcal{C}_2\mathcal{D}_2-4\mathcal{B}_2 \mathcal{C}_2}{(2r_h^2-\gamma \xi^2)\mathcal{B}_2^2}.
\end{split}
\end{eqnarray}

Within the parameter range considered here, the surface gravity remains strictly positive, i.e.$f^\prime_h > 0$, which corresponds to the non-extremal horizon case. The surface gravity and Hawking temperature of the black hole are given by
\begin{eqnarray}\label{eq. Hawking}
\kappa&=&\frac{1}{2}f^\prime_h=\frac{r_h\mathcal{C}_2\left[r_h(\xi^2+\mathcal{A}_2)-2\mathcal{B}_2\right]}{(2r_h^2-\gamma \xi^2)\mathcal{B}_2^2}\,,\\
T_H&=&\frac{r_h\mathcal{C}_2\left[r_h(\xi^2+\mathcal{A}_2)-2\mathcal{B}_2\right]}{2\pi(2r_h^2-\gamma \xi^2)\mathcal{B}_2^2}.
\end{eqnarray}

\begin{figure*}[!ht]
\includegraphics[width=0.4\textwidth]{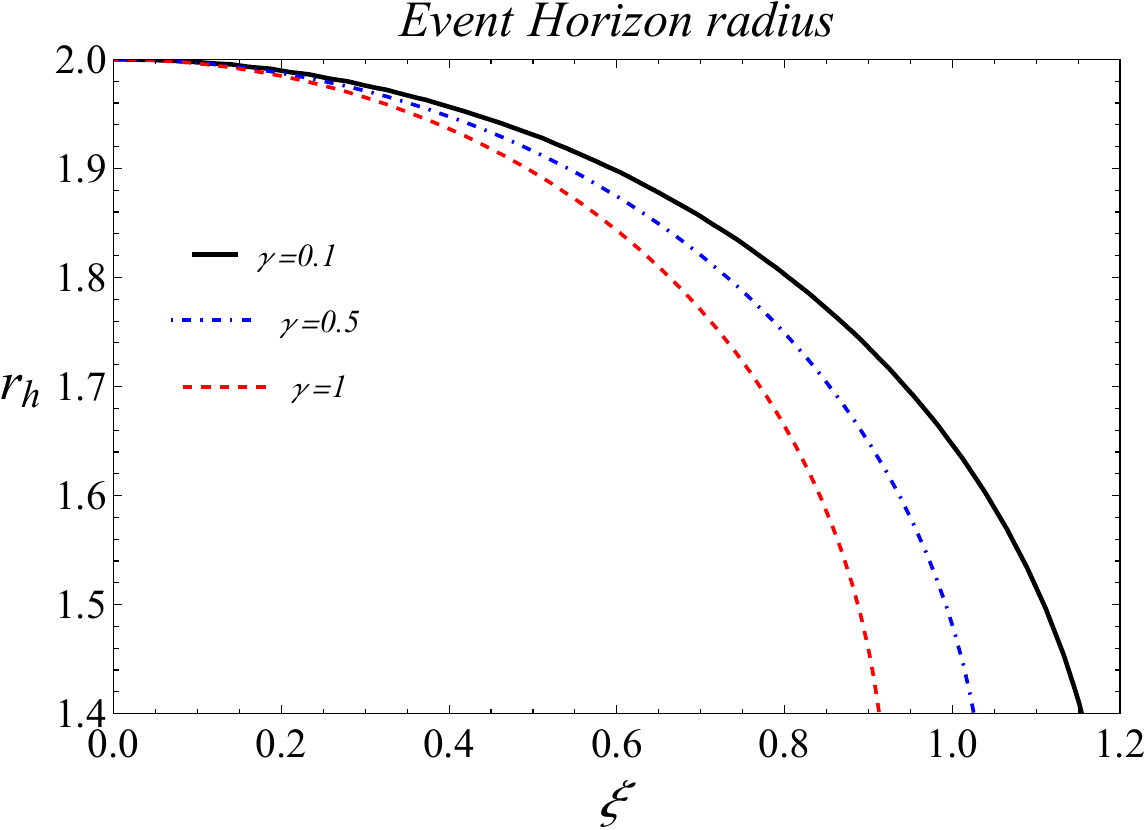}
\includegraphics[width=0.4\textwidth]{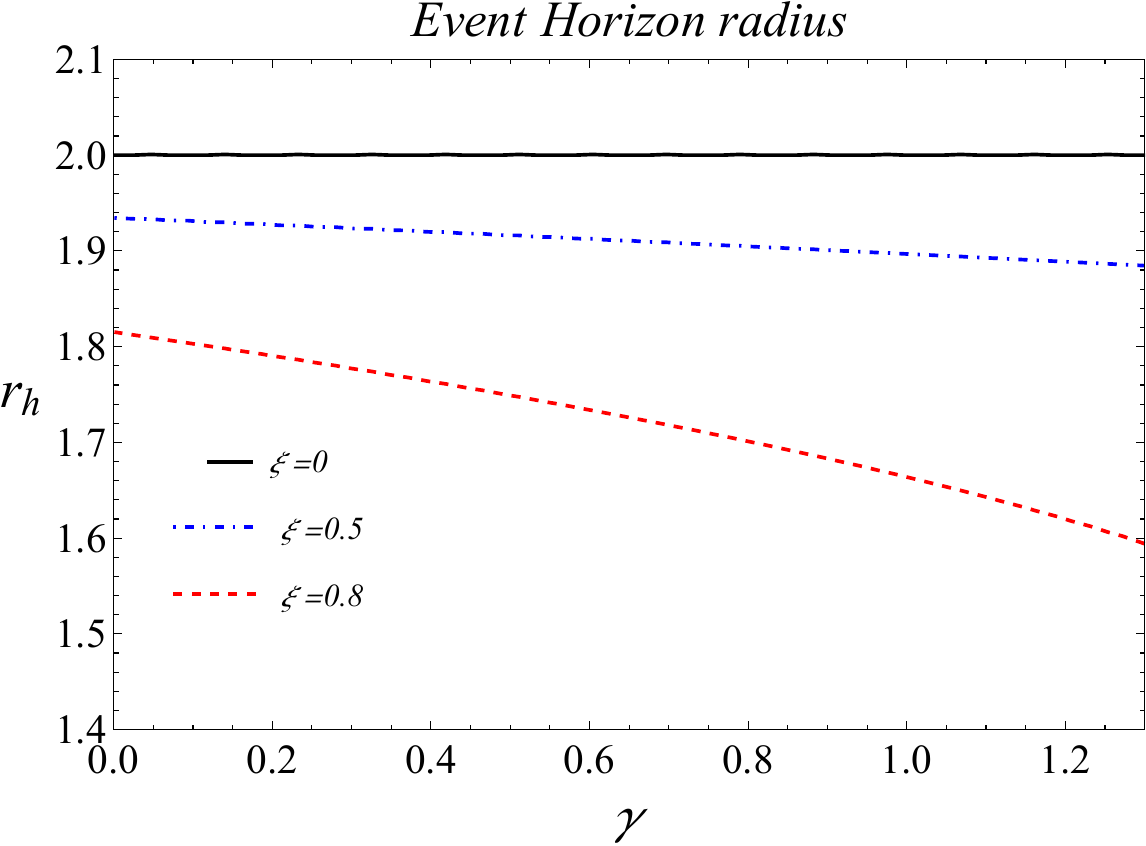}
\caption{Variation of the horizon radius $r_h$ with the quantum parameter $\xi$.\label{Fig.r_h}}
\end{figure*}

\begin{figure*}[!ht]
\includegraphics[width=0.4\textwidth]{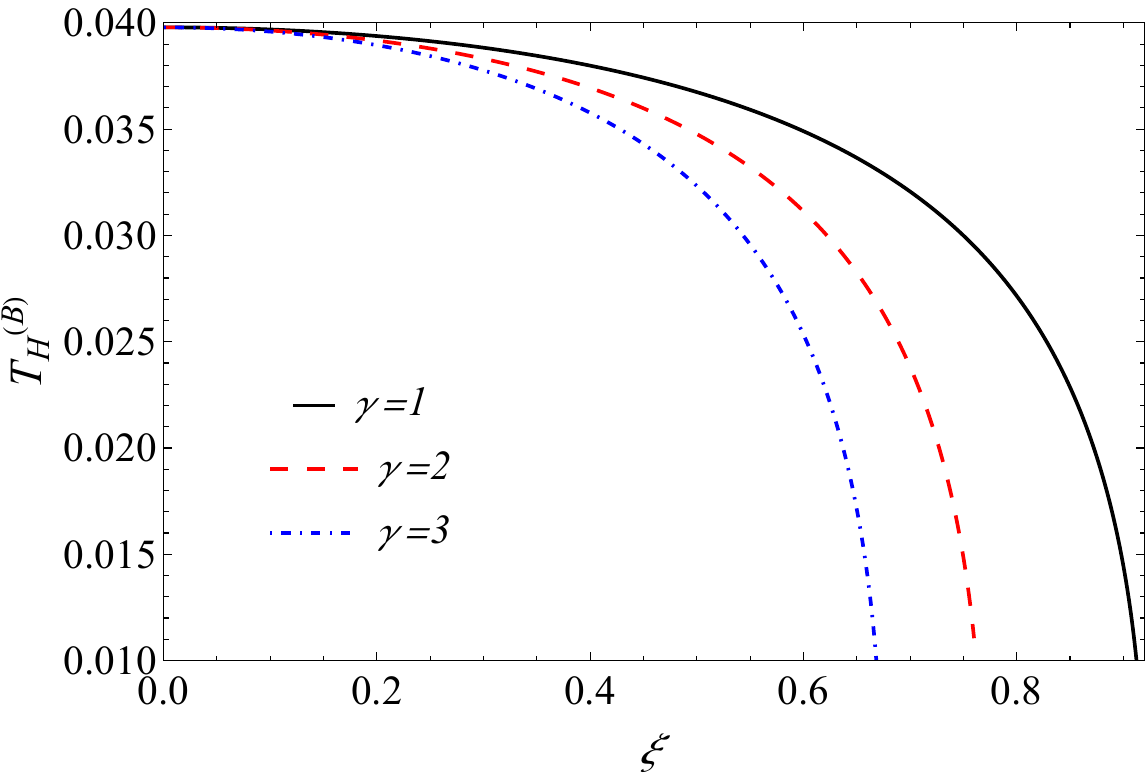}
\includegraphics[width=0.4\textwidth]{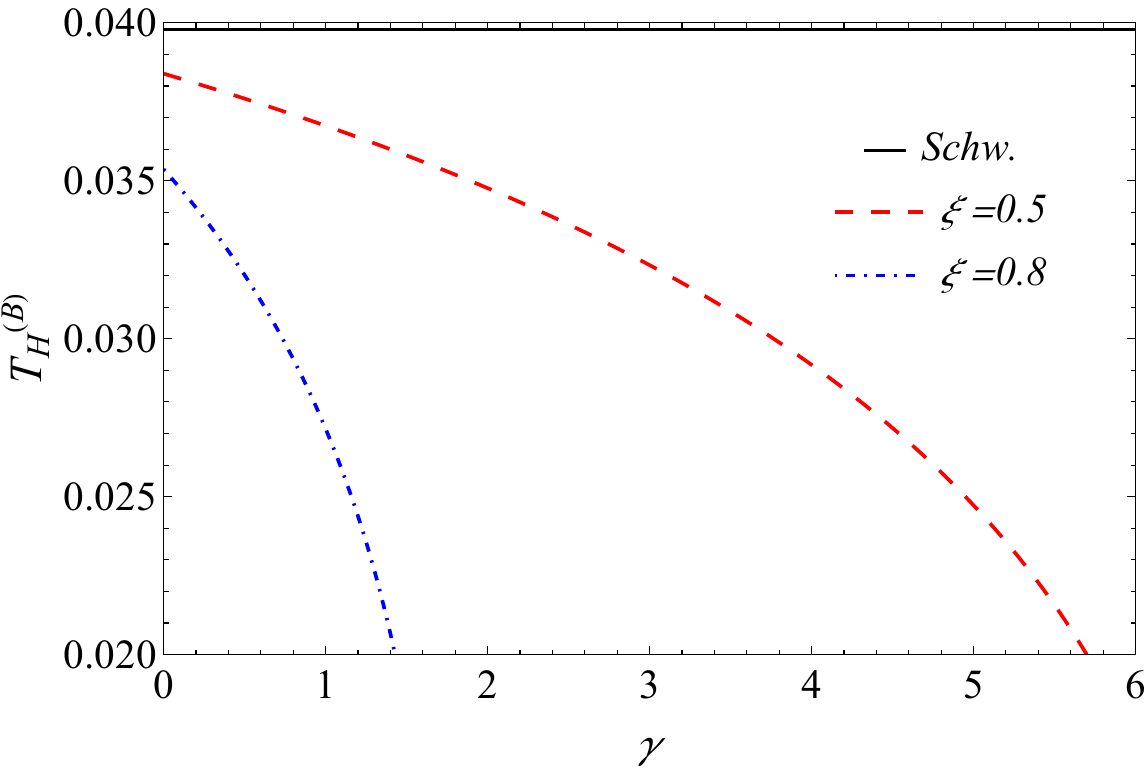}
\caption{Dependence of the Hawking temperature $T_H$ on the parameter $\gamma$ and $\xi$ for different solutions.\label{Fig.TH}}
\end{figure*}

Figure 1 shows the dependence of the radius of the horizon $r_h$ on the parameter $\gamma$ and $\xi$, while Figure 2 shows the dependence of the Hawking temperature $T_H$ on $\gamma$ and $\xi$. One can see from Figs.(\ref{Fig.r_h},\ref{Fig.TH}) that increasing both quantum correction parameters $\xi$ and $\gamma$ causes shrinking event horizon and decreasing the value of the Hawking temperature $T_H$.

In what follows, we will analyze the thermal spectral characteristics of the atomic excitation probability and investigate the influence of the quantum correction parameter on acceleration radiation within this framework.

\section{Geodesic equations}\label{Sec.II}

To describe the radial infall of an atom released from rest at infinity, we solve the geodesic equations in a static, spherically symmetric spacetime (\cite{UktamjonUktamov:2025aqz,Uktamjon:2024fjb,UktamjonUktamov:2025emm,Uktamov:2025bth}):
\begin{eqnarray}\label{eq. Geodesic}
\frac{d^2x^\mu}{d\tau^2}+\Gamma^\mu_{\rho \sigma}\frac{dx^\rho}{\tau}\frac{dx^\sigma}{d\tau}=0,
\end{eqnarray}
where $\tau$ is the proper time and are $\Gamma^\mu_{\rho \sigma}$ the Christoffel symbols associated. 

For metric (\ref{eq.metric}), spherical symmetry allows us to restrict the motion to the equatorial plane $(\theta=\frac{\pi}{2})$ and set the angular momentum to zero$(\dot{\theta}=\dot{\phi}=0)$, resulting in purely radial motion.The conserved energy per unit mass is
\begin{eqnarray}\label{eq. energy}
\mathcal{E} = \left[1-\frac{4M r^2}{\xi^2(\gamma M+r)+\sqrt{\xi^4(\gamma M+r)^2+4r^6}}\right]\frac{dt}{d\tau}.
\end{eqnarray}
The four-velocity $u^\mu=dx^\mu/d\tau$ satisfies the normalization $u^\mu u_\mu=-1$, yielding
\begin{eqnarray}\label{eq. motion}
\left(\frac{dr}{d\tau}\right)^2&=&\mathcal{E}^2-f(r),\\
\left(\frac{dr}{dt}\right)^2&=&\frac{f(r)^2}{\mathcal{E}^2}\left[\mathcal{E}^2-f(r)\right].
\end{eqnarray}

Using the radial velocity relations derived above, we integrate the equations of motion with respect to the radial coordinate to obtain the proper time $\tau$ and coordinate time $t$ elapsed as the atom travels from an initial radius $r_i$ to a final radius $r_f$:
\begin{eqnarray}\label{eq. time}
\tau&=&-\int_{r_i}^{r_f}\frac{dr}{\sqrt{\mathcal{E}^2-f(r)}}\,,\\ \label{eq. time2}
t&=&-\int_{r_i}^{r_f}\frac{\mathcal{E}}{f(r)\sqrt{\mathcal{E}^2-f(r)}}dr,
\end{eqnarray}
for an atom released from rest at infinity, the conserved energy takes the value $\mathcal{E}=1$. The minus signs in the above expressions indicate radial infall, i.e., the radial coordinate $r$ decreases as the proper time $\tau$ and coordinate time $t$ increase.

Near the horizon $r=r_h$, we introduce the shifted coordinate $x=r-r_h$.  For a non-extremal black hole, the metric function expands as $f(r)\approx f^\prime_hx+\mathcal{O}(x^2)$. Accordingly, the near-horizon expansions of $f(r)$ and its derivatives can be organized as a hierarchical series in powers of $x$.

\begin{eqnarray}\label{eq. expand1}
f(r)&\overset{r_h}{\sim}&f^\prime_h x\left[1+\mathcal{O}(x)\right]\,,\\\label{eq. expand2}
f^\prime(r)&\overset{r_h}{\sim}&f^\prime_h \left[1+\mathcal{O}(x)\right]\,,\\\label{eq. expand3}
f^{\prime\prime}(r)&\overset{r_h}{\sim}&f^{\prime\prime}_h \left[1+\mathcal{O}(x)\right],
\end{eqnarray}
where $f^\prime_h$ and $f^{\prime\prime}_h$ are given in Eqs.(\ref{eq. derivatives rh},\ref{eq. derivatives rh2}) and the symbol denotes the $\overset{r_h}{\sim}$ denotes an expansion performed near the horizon.For a non-extremal black hole, the metric function behaves linearly near the horizon as $f(r)\approx f^\prime_hx$.

Substituting this expansion into the integral expressions Eqs.(\ref{eq. time},\ref{eq. time2}) for the proper time and coordinate time, and expanding the integrand $\sqrt{\mathcal{E}^2-f(r)}$ in powers of $x$, we keep only the leading-order terms. This yields the near-horizon approximations:
\begin{eqnarray}\label{eq. time3}
\tau&=&-\frac{1}{\mathcal{E}}\left(x+\frac{f^\prime_h}{4\mathcal{E}^2}x^2\right)+C\,,\\\label{eq. time4}
t&=&-\frac{1}{f^\prime_h}lnx-\frac{1}{2\mathcal{E}^2}x+\frac{f^{\prime \prime}_h}{2f^{\prime 2}_h}x+\frac{f^{\prime \prime}_h}{8\mathcal{E}^2 f^\prime_h}x^2+C.
\end{eqnarray}

Thus, the proper time to reach the horizon is finite, while the coordinate time diverges logarithmically, reflecting the infinite redshift as seen by a distant observer. These expansions serve as the basis for the detector response calculation in later sections.

\section{Near-horizon scalar field modes and conformal quantum mechanics reduction}\label{Sec.III}
To further investigate the interaction between the atom and the quantum field, in this section we establish the dynamics of the scalar field in the near-horizon region of this background. We consider a massless scalar field, as it is sufficient to reveal the thermal characteristics of the HBAR spectrum and is fully compatible with the conformal quantum mechanics framework. The dynamics of this field is governed by the Klein–Gordon equation (\cite{Pantig:2025okn}):
\begin{eqnarray}\label{eq. field}
\frac{1}{\sqrt{-g}}\partial_\mu\left(\sqrt{-g}g^{\mu\nu}\partial_\nu\Phi\right)=0\,,
\end{eqnarray}
where $g^{\mu\nu}$ is the  metric given by Eq.(\ref{eq.metric}). Using spherical symmetry, the field can be expanded in spherical harmonics. For the HBAR process, however, the dominant contribution comes from the s-wave ($l=0$) modes, and the angular barrier is negligible near the horizon. Hence we retain only the s-wave component and define the reduced field $\Psi(t,r)=r\Phi(t,r)$. Substituting into the metric, the Klein–Gordon equation reduces to the following ($1+1$)-dimensional form:
\begin{eqnarray}\label{eq. field2}
-\frac{1}{f(r)}\frac{\partial^2\Psi}{\partial t^2}+\frac{\partial}{\partial r}\left[f(r)\frac{\partial \Psi}{\partial r}\right]=0\,.
\end{eqnarray}

In this equation, the redshift function $f(r)$ encodes all the effects of the background geometry on the field propagation. Near the horizon, $f(r)\to0$, so that time derivative term is strongly amplified while the radial derivative term is suppressed.

We expand the field in modes as (\cite{Jana:2024fhx}):

\begin{eqnarray}\label{eq. field2}
\Psi(t,r)&=&\sum[\hat{a}\psi(r,t)+H.c.]\,,\\\nonumber
\psi(t,r)&=&\chi(r)u(r)e^{-i\nu t}
\end{eqnarray}
to eliminate the first-derivative term, we perform a standard transformation as $\chi(r)=[f(r)]^{-\frac{1}{2}}\,$, then we have the following Schrödinger-like radial equation:
\begin{eqnarray}\label{eq. u}
\frac{d^2u}{dr^2}+V_{eff}(r;\nu)u=0\,,
\end{eqnarray}
where the effective potential $V_{eff}(r;\nu)$ is given by:
\begin{eqnarray}\label{eq. veff}
V_{eff}(r;\nu)=\frac{\nu^2}{f(r)^2}+\frac{f^\prime(r)^2}{4f(r)^2}-\frac{f^{\prime\prime}(r)}{2f(r)}\,,
\end{eqnarray}

Using the near-horizon expansions Eqs.(\ref{eq. expand1}-\ref{eq. expand3}) of the redshift function to simplify, and ignoring higher-order terms, we can obtain the radial equation of the near event horizon in CQM form:
\begin{eqnarray}\label{eq. CQM}
\frac{d^2u}{dr^2}+\frac{\lambda_{eff}}{x^2}[1+\mathcal{O}(x)]u=0\,,
\end{eqnarray}
where,
\begin{eqnarray}\label{eq. lambda}
\lambda_{eff}=\frac{1}{4}+\Theta^2\,,\quad \Theta=\frac{\nu}{f^\prime_h}.
\end{eqnarray}

The independent solutions are of power-law form $u(x)\propto x^\alpha$. Substituting gives the indicial equation $\alpha(\alpha-1)+\lambda_{eff}=0$, whose solutions are:
\begin{eqnarray}\label{eq. alpha}
\alpha=\frac{1}{2}\pm i\Theta\,,
\end{eqnarray}
the two solutions correspond to ingoing and outgoing waves. To describe radiation propagating outward from the horizon, we select the outgoing mode, taking the sign that yields an outgoing flux. Explicitly, we choose
\begin{eqnarray}\label{eq. alpha}
u_{our}(x)=Cx^{\frac{1}{2}+i\Theta}\,,
\end{eqnarray}
where C is the normalization constant, combining with the transformation factor $\chi$ and the time factor $e^{-i\nu t}$, we obtain the full mode function in the near-horizon region:
\begin{eqnarray}\label{eq. phi}
\phi(t,r)=\frac{1}{\sqrt{f^\prime_h}}x^{i\Theta}e^{-i\nu t}\,,
\end{eqnarray}
where we have absorbed constants. This expression can be rewritten more compactly using the tortoise coordinate $r_*=\int dr/f(r)\approx(1/f^\prime_h)lnx+\mathcal{O}(x)$
\begin{eqnarray}\label{eq. phi tortoise}
\phi(t,r)=\frac{1}{\sqrt{f^\prime_h}}e^{-i\nu (t-r_*)}\,,
\end{eqnarray}
this form has been widely used in HBAR theory and will be directly employed in the next section to compute the atomic excitation probability.

\section{Atomic excitation probability and the HBAR spectrum}\label{Sec.IV}

To compute the interaction between the freely falling atom and the scalar field, we follow the model in (\cite{UktamjonUktamov:2026dep}), which consists of mirrors surrounding a black hole to avoid the effects of Hawking radiation. To compute the interaction between the freely falling atom and the quantum field, we adopt the standard Unruh–DeWitt detector model. The atom has two energy levels: ground state $\vert b\rangle$ and excited state $\vert a\rangle$. The atom moves along the radial geodesic, parameterized by its proper time $\tau$. The field is prepared in the Boulware vacuum $\vert 0\rangle$, which is regular at the horizon and contains no outgoing Hawking quanta. In the presence of a stretched mirror, we consider a single field mode of frequency $\nu$. The initial state of the combined system is $\vert 0,b\rangle$, and the final state is $\vert 1_\nu,b\rangle$. By first-order time-dependent perturbation theory, the excitation probability is:
\begin{eqnarray}\label{eq. probability}
P_{exc}=\left\vert\int\langle1_\nu,a\vert V(\tau)\vert 0,b\rangle d\tau\right\vert^2\,,
\end{eqnarray}
where $\tau$ is the proper time along the atomic trajectory. The interaction Hamiltonian in the interaction picture can be taken as (\cite{Ovgun:2025isv}):
\begin{eqnarray}\label{eq. Hamiltonian}
\begin{split}
V(\tau)=&g_c\left[\hat{a}\phi\left(r(\tau),t(\tau)\right)+\hat{a}^\dagger\phi^*\left(r(\tau),t(\tau)\right)\right]\\\times&(\sigma_{-} e^{-i\omega\tau}+H.c.)\,,
\end{split}
\end{eqnarray}
where $\sigma_{-}$ is the atomic lowering operator, $g_c$ is the atom-field coupling constant, and $\omega$ is the atomic transition frequency. Since $\sigma_{-}$ vanishes when acting on the ground state and only the creation part of the field operator (proportional to $\phi^*$) can promote the field from vacuum to a single-particle state, only the $\hat{a}^\dagger\sigma_{+}$ term contributes. Substituting the Hamiltonian into Eq.(\ref{eq. probability}), we can obtain:
\begin{eqnarray}\label{eq. probability2}
P_{exc}=g_c^2\left\vert\int\phi^*\left(r(\tau),t(\tau)\right)e^{-i\omega\tau} d\tau\right\vert^2\,.
\end{eqnarray}

This expression indicates that the excitation probability is determined by the coherent overlap between the atom transition frequency and the phase evolution of the field mode along the atomic trajectory. Considering the near-horizon outgoing mode function Eqs.(\ref{eq. phi},\ref{eq. phi tortoise}) and the near-horizon geodesic expansion Eqs.(\ref{eq. time3},\ref{eq. time4})
\begin{eqnarray}\label{eq. phi2}
\phi(t,r)&=&\frac{1}{\sqrt{f^\prime_h}}e^{-i\nu (t-r_*)}=\frac{1}{\sqrt{f^\prime_h}}x^{i\Theta}e^{-i\nu t}\,,
\end{eqnarray}
ignoring constant offsets, then:
\begin{eqnarray}\label{eq. phi horizen}
P_{exc}=g_c^2\left\vert\int_{0}^{x_f}x^{-2i\Theta}e^{-i\omega x} dx\right\vert^2\,.
\end{eqnarray}
where $x_f$ is the upper cutoff of the radial integral, satisfying $x_f\ll r_h$ and being larger than the minimal scale required for the validity of the near-horizon expansion. This cutoff reflects the fact that our near-horizon expansions are valid only in the region $x\leq x_f$. 

However, the integral receives its dominant contribution from $x\sim 1/\omega$. In the geometric optics limit, where the atomic transition frequency $\omega$ is much larger than any other relevant scale, we have $1/\omega \ll x_f$. This implies that the integrand oscillates rapidly in the interval $x\leq 1/\omega$, while for $x\gg1/\omega$ the oscillations cause cancellations that suppress the contribution. Consequently, extending the upper limit from $x_f$ to infinity does not alter the leading behavior of the integral.  Thus we may write the integral as $\int_0^\infty x^{-2i\Theta}e^{-i\omega x} dx$ and evaluate it using standard Gamma-function formulas.
\begin{eqnarray}\label{eq. inf}
\int_0^\infty x^{2i\nu}e^{ix} dx=-\frac{xe^{-\pi \nu}}{sinh(2\pi\nu)\Gamma(-2i\nu)}\,,
\end{eqnarray}
with a correspondence of exponents $2i\nu\to-2i\Theta$ and argument rescaling $x\to-\omega x$, its modulus squared is:
\begin{eqnarray}\label{eq. mode square}
\left\vert\int_{0}^{\infty}x^{-2i\Theta}e^{-i\omega x} dx\right\vert^2=\frac{4\pi\Theta}{\omega^2}\frac{1}{e^{4\pi\Theta}-1}\,,
\end{eqnarray}
Substituting the squared modulus into the excitation probability formula yields
\begin{eqnarray}\label{eq. probability2}
P_{exc}=\frac{4\pi g_c^2\Theta}{\omega^2}\frac{1}{e^{4\pi\Theta}-1}\,,
\end{eqnarray}
it is convenient to rewrite the result in terms of the surface gravity $\kappa$ and  Hawking temperature $T_H$: 
\begin{eqnarray}\label{eq. probability3}
P_{exc}=\frac{2\pi g_c^2\nu}{\kappa\omega^2}\frac{1}{exp\left(\frac{\nu}{T_H}\right)-1}\,.
\end{eqnarray}
The excitation probability (\ref{eq. probability3}) contains a Planck factor with temperature equal to the  Hawking temperature $T_H$. This shows that even for a quantum-corrected BH (with no central singularity), as long as the horizon is non-extremal, the radiation detected by a freely falling atom is thermal, with the temperature determined by the surface gravity.

The parameters $\gamma$ and $\xi$ affects the excitation probability through the surface gravity $\kappa$ and hence through $T_H$ as well as the frequency $\nu$. In the extremal limit $\kappa\to0$ and thus $T_H\to0$, the Planck factor is exponentially suppressed, and the excitation probability tends to zero.

\begin{figure*}[ht!]
\includegraphics[width=0.45\textwidth]{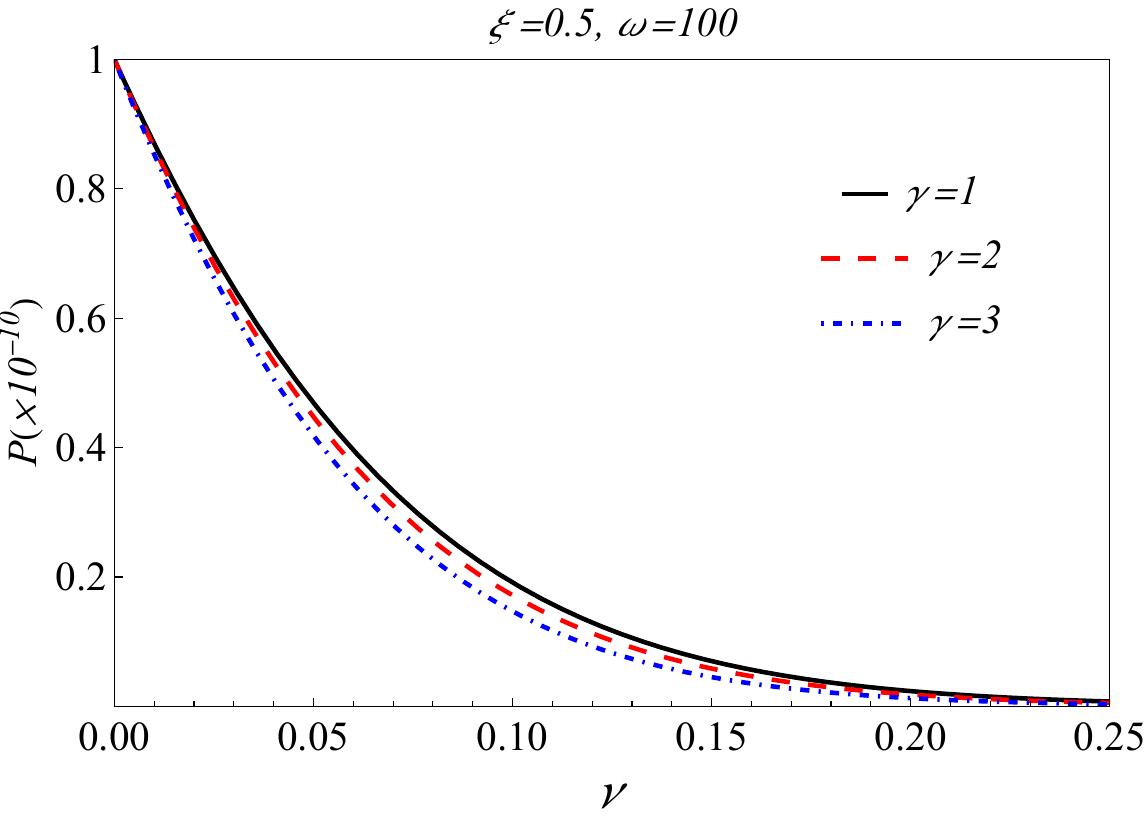}
\includegraphics[width=0.45\textwidth]{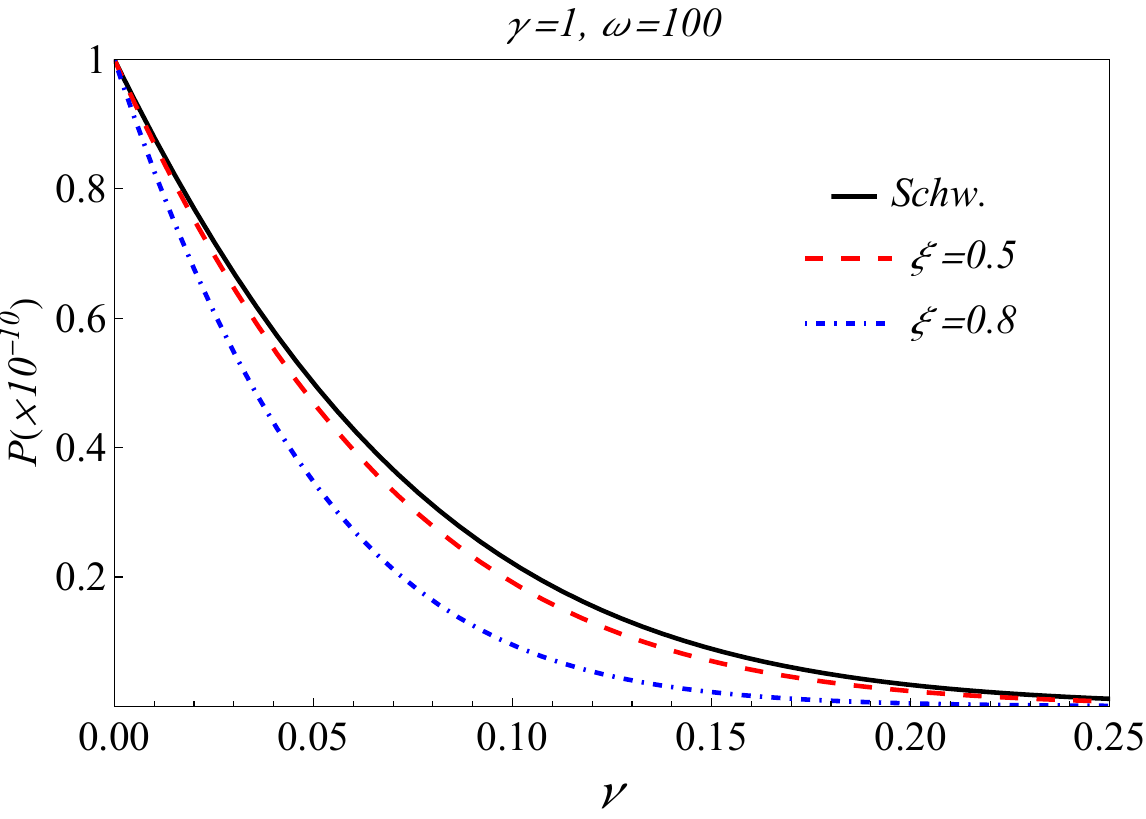}
\includegraphics[width=0.45\textwidth]{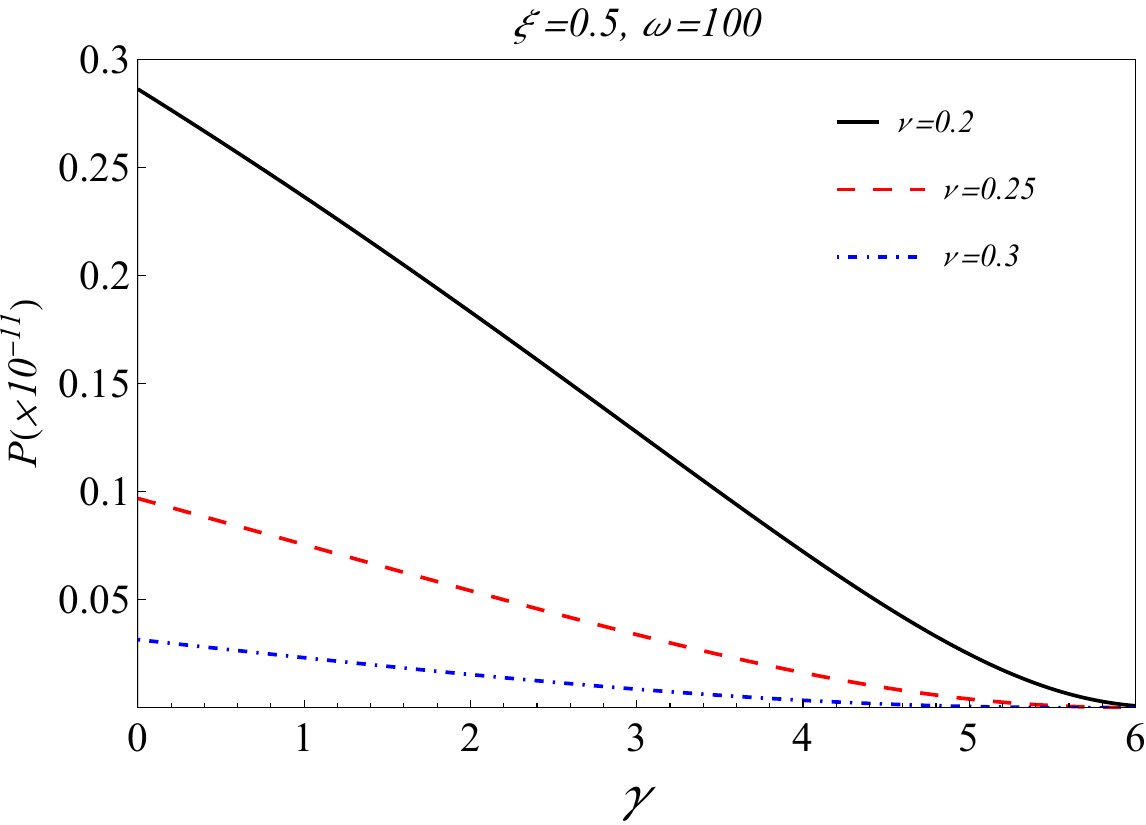}
\includegraphics[width=0.45\textwidth]{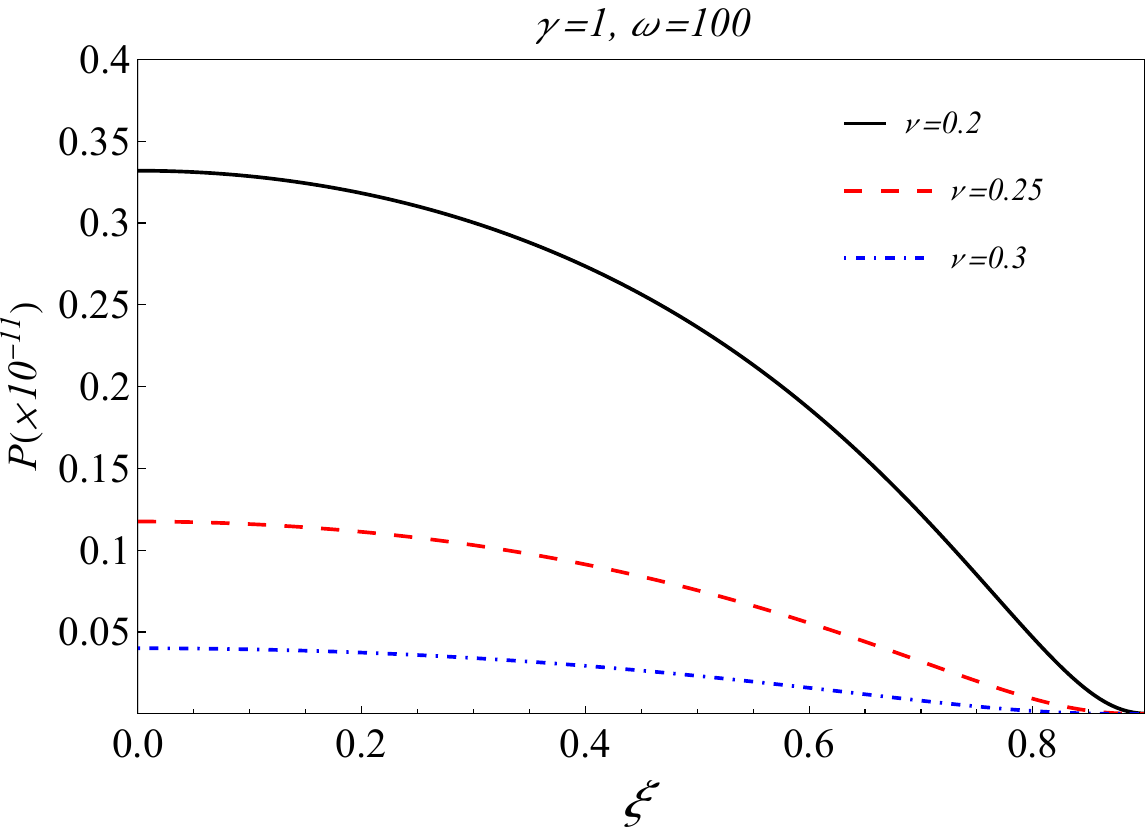}
\includegraphics[width=0.45\textwidth]{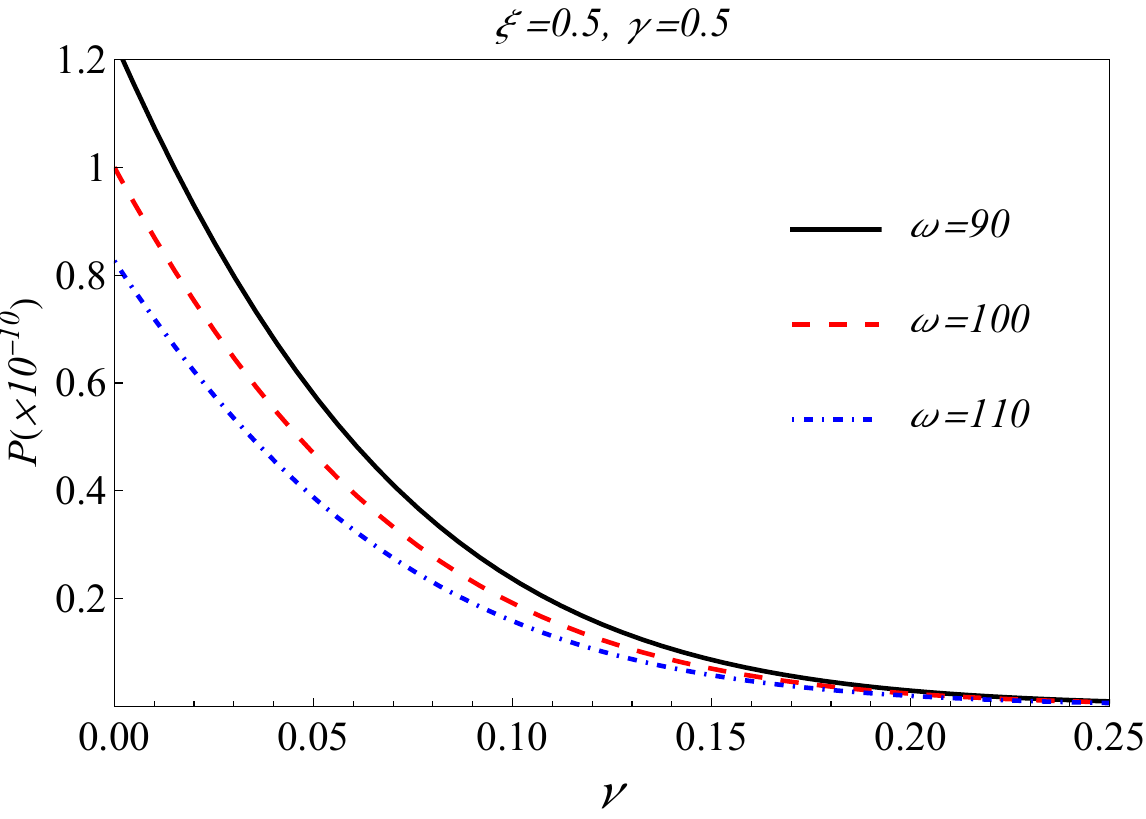}
\includegraphics[width=0.45\textwidth]{P-niu_different_xi.pdf}
\caption{\label{Fig.Probability}Variation of the excitation probability $P_{exc}$ with different parameters. (The first line) $P_{exc}$ as a function of mode frequency $\nu$ for different values of metric parameters($\gamma$ and $\xi$). (The second line)$P_{exc}$ as a function of metric parameters for different values of $\nu$. (The third line left) $P_{exc}$ as a function of mode frequency $\nu$ for different atomic transition frequencies $\omega$, (The third line right) Probability difference $\Delta P$ as a function of $\nu$ for different values of $\xi$.}
\end{figure*}

Fig ~\ref{Fig.Probability} shows the dependence of the excitation probability $P_{exc}$ on the mode frequency $\nu$, the atomic transition frequency $\omega$, and the metric parameters $\gamma$ and $\xi$. The ADM mass is fixed to $M=1,$ and the coupling constant is $g_c=10^{-3}$, working in the geometric optics regime ($\omega\gg\nu$).

First line for different values of the parameter $\gamma$ ($\gamma=1, 2, 3$) and $\xi$ (from the Schwarzschild limit $\xi=0$ to nearly extremal), $P_{exc}$ is plotted against $\nu$. Each curve exhibits a Planck-type spectrum: the probability tends to a constant at low frequencies and decays exponentially at high frequencies. As the parameters increases, the surface gravity $\kappa$ decreases, and the Hawking temperature $T^{(B)}_H$ decreases accordingly. The entire spectrum shifts toward lower frequencies, and its amplitude drops significantly. This directly reflects the “cooling” effect of the regular core on the HBAR signal.

Second line fixing $\omega=100$, $P_{exc}$ is plotted as a function of the parameters $\gamma$ and $\xi$ for several representative mode frequencies $\nu$ ($\nu=0.2, 0.25, 0.3$). For all $\nu$, the probability decreases monotonically with increasing $\gamma$ and $\xi$ and tends to zero as the parameters approaches the extremal limit. This confirms the tunable suppression of acceleration radiation by the regular core parameter.

On the left of the third line, for the fixed parameters ($\xi=0.5,\gamma=0.5$) and varying atomic transition frequency $\omega$ ($\omega=90, 100, 110$), the probability amplitude decreases as $\omega^{-2}$, while the characteristic temperature of the Planck factor remains unchanged. This indicates that the atomic energy gap affects only the signal strength, not the thermal shape. On the right, the difference $\Delta P=P_{exc}^{Schw}-P_{exc}$ is shown, i.e., the reduction in excitation probability of the quantum corrected black hole relative to the Schwarzschild case. This difference is largest at low frequencies and decays with increasing $\nu$. Moreover, the larger the parameter $\xi$, the larger the difference, indicating that the suppression of HBAR becomes stronger as the parameter increases.

\begin{figure*}[ht!]
\includegraphics[width=0.45\textwidth]{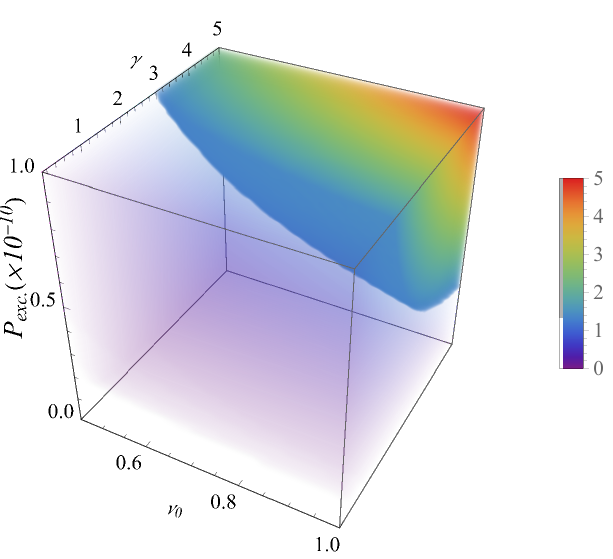}
\includegraphics[width=0.45\textwidth]{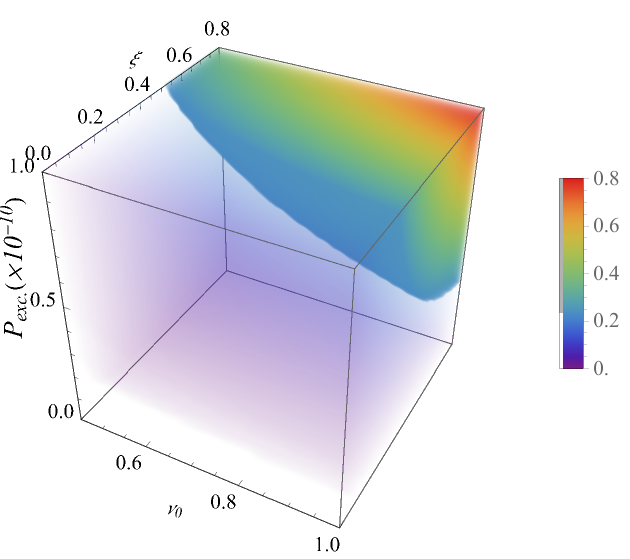}
\caption{\label{Fig.3D}Three-dimensional view of the excitation probability $P_{exc}$ as a function of frequency $\nu_0$ and the parameters $\gamma$ and $\xi$.}
\end{figure*}

Fig. \ref{Fig.3D} displays the overall variation of the excitation probability $P_{exc}$ with frequency and the parameters $\gamma$ and $\xi$ in a three-dimensional view. It is evident that the largest excitation probabilities are concentrated in the region of low frequency and small parameter values, while the probability drops rapidly as either frequency or parameters increase. Along the frequency direction, slices at fixed parameters exhibit a typical Planckian profile, whose peak shifts to lower frequencies and decreases in magnitude with growing parameters. Along the parameter direction, for a fixed frequency, the probability decreases monotonically and tends to zero when the parameters approach their critical values. This three-dimensional representation clearly illustrates the dependence of the excitation probability on the entire parameter space.

\section{HBAR entropy and WIEN's displacement law}\label{Sec.V}

To embed the microscopic transition probabilities into a macroscopic thermodynamic framework, we adopt the standard open-system approach in quantum optics. Let $\rho$ be the density matrix of the radiation field mode, whose evolution is governed by the Lindblad master equation,
\begin{eqnarray}\label{eq. rho}
\dot{\rho}=-i[H_{eff},\rho]+\Gamma_{em}\mathcal{D}[a^\dagger]\rho+\Gamma_{abs}\mathcal{D}[a]\rho\,,
\end{eqnarray}
where $H_{eff}=\nu_0a^\dagger a$ is the effective Hamiltonian, the dissipator is defined as $\mathcal{D}[O]\rho=O\rho O^\dagger-\frac{1}{2}\left\{O^\dagger O,\rho\right\}$.
With the continuous supply of two-level atoms, each possessing an energy gap $\omega$, into the cavity at a flux $\mathcal{I}$, the emission and absorption rates, denoted by $\Gamma_{em}$ and $\Gamma_{abs}$, are respectively expressed as
\begin{eqnarray}\label{eq. rate}
\Gamma_{em}&=&\mathcal{I}P_{exc}\quad\Gamma_{abs}=\mathcal{I}P_{abs}\,,\\
\frac{\Gamma_{em}}{\Gamma_{abs}}&=&e^{-\frac{\nu_0}{T_{H}^{(B)}}}\,.
\end{eqnarray}
Taking the Fock-state diagonal elements $\rho_{n,n}\equiv\langle n\vert \rho\vert n\rangle$, the master equation yields the rate equation for the photon-number probabilities:
\begin{eqnarray}\label{eq. rho nn}
\begin{split}
\dot{\rho}_{n,n}=&\Gamma_{em}\left[n\rho_{n-1,n-1}-(n+1)\rho_{n,n}\right]\\&+\Gamma_{abs}\left[(n+1)\rho_{n+1,n+1}-n\rho_{n,n}\right]\,.
\end{split}
\end{eqnarray}
Under steady-state conditions ($\dot{\rho}_{n,n}=0$), combined with the detailed-balance ratio, we immediately obtain the ratio of adjacent occupation numbers:
\begin{eqnarray}\label{eq. rhorho}
\frac{\rho_{n,n}}{\rho_{n-1,n-1}}=\frac{\Gamma_{em}}{\Gamma_{abs}}=e^{-\frac{\nu_0}{T_H}^{(B)}}\,,
\end{eqnarray}
Solving iteratively and imposing the normalization $\sum_n\rho_{n,n}=1$, we arrive at the steady-state Bose–Einstein distribution:
\begin{eqnarray}\label{eq. rhos}
\rho_{n,n}^{(s)}=\left(1-e^{-\nu_0/T_H^{(B)}}\right)e^{-n\nu_0/T_H^{(B)}}\,,
\end{eqnarray}
with the corresponding mean photon number
\begin{eqnarray}\label{eq. n}
\bar{n}_{\nu_0}=\sum_nn\rho_{n,n}^{(s)}=\left(e^{\nu_0/T_H^{(B)}}-1\right)^{-1}\,,
\end{eqnarray}

Furthermore, the total field energy and its rate of change are
\begin{eqnarray}\label{eq. field energy}
E_P=\sum_{\nu_0}\nu_0\bar{n}_{\nu_0}\quad\dot{E}_P=\sum_{\nu_0}\nu_0\dot{\bar{n}}_{\nu_0}\,,
\end{eqnarray}
In the dilute limit ($\bar{n}_{\nu_0}\ll1$), the the mode occupation number gives 
\begin{eqnarray}\label{eq. ndot}
\dot{\bar{n}}_{\nu_0}=\Gamma_{em}(\bar{n}_{\nu_0}+1)-\Gamma_{abs}\bar{n}_{\nu_0}\simeq\Gamma_{em}=\mathcal{I}P_{exc}\,,
\end{eqnarray}
Substituting Eq.(\ref{eq. field energy},\ref{eq. ndot}) into Eq.(\ref{eq. probability3}), we obtain the HBAR energy flux:
\begin{eqnarray}\label{eq. energydot}
\dot{E}_P\simeq\frac{g_c^2\mathcal{I}}{\omega^2}\sum_{\nu_0}\frac{\nu_0^2}{T_H^{(B)}}\frac{1}{exp\left(\frac{\nu}{T^{(B)}_H}\right)-1}\,.
\end{eqnarray}
On the other hand, the von Neumann entropy of the radiation field is
\begin{eqnarray}\label{eq. S}
S_\rho=\sum_{n,\nu}\rho_{n,n}ln\rho_{n,n}\,,
\end{eqnarray}
and its rate of change resulting from photon emission is
\begin{eqnarray}\label{eq. Sdot}
\dot{S_\rho}=-\sum_{n,\nu}\dot{\rho}_{n,n}ln\rho_{n,n}^{(s)}\,.
\end{eqnarray}
Inserting the steady-state distribution Eq.(\ref{eq. rhos}) and using the conservation of trace $\sum_n\dot{\rho}_{n,n}=0$, the relationship between HBAR entropy and energy flux:
\begin{eqnarray}\label{eq. entropy}
\dot{S_\rho}=\frac{\dot{E}_P}{T_H^{(B)}}=\frac{2\pi(2r_h^2-\gamma \xi^2)\mathcal{B}_2^2}{r_h\mathcal{C}_2\left[r_h(\xi^2+\mathcal{A}_2)-2\mathcal{B}_2\right]}\dot{E}_P\,,
\end{eqnarray}
which is precisely the Clausius-type first law $\dot{S}=\dot{E}/T$ realized in the quantum-corrected black hole background.

To establish an observable diagnostic directly linked to the preceding thermodynamic properties, we now transform the frequency-domain HBAR excitation spectrum into the wavelength domain and determine its peak-shift behavior. Define the wavelength-spectral density $P_\lambda(\lambda)$ to satisfy
\begin{eqnarray}\label{eq. Plambda}
P_\lambda(\lambda)d\lambda=P_{\nu_0}(\nu_0)d\nu_0\quad\nu_0=1/\lambda\,,
\end{eqnarray}
we obtain from Eq.(\ref{eq. probability3}):
\begin{eqnarray}\label{eq. Pro lambda}
P_\lambda(\lambda)\propto\frac{1}{\lambda^3}\frac{1}{exp\left(1/\lambda T_H^{(B)}\right)-1}\,.
\end{eqnarray}
The peak wavelength $\lambda_{crit}$ is fixed by the extremum condition $dP_\lambda(\lambda)/d\lambda|_{\lambda=\lambda_{crit}}=0$. Introducing the dimensionless variable $x=1/(\lambda T_H^{(B)})$, the extremum condition reduces to the universal equation in terms of $x$:
\begin{eqnarray}\label{eq. x}
\frac{d}{dx}\left[\frac{x^3}{e^x-1}\right]_{x=x_{crit}}=0\,,
\end{eqnarray}
which is equivalent to $1-e^{-x_{crit}}=x_{crit}/3$. This transcendental equation is independent of the specific black hole parameters, and its unique positive root is $x_{crit}\simeq2.82144$. Restoring SI units with $\hbar$ and $k_B$, the peak wavelength becomes:
\begin{eqnarray}\label{eq. lambda crit}
\begin{split}
\lambda_{crit}^{(B)}&\simeq\frac{2\pi\hbar}{x_{crit}k_BT_H^{(B)}}\approx2.23\frac{\hbar}{k_BT_H^{(B)}}\\
&=2.23\frac{h(2r_h^2-\gamma \xi^2)\mathcal{B}_2^2}{k_Br_h\mathcal{C}_2\left[r_h(\xi^2+\mathcal{A}_2)-2\mathcal{B}_2\right]}\,,
\end{split}
\end{eqnarray}

\begin{figure*}[ht!]
\includegraphics[width=0.45\textwidth]{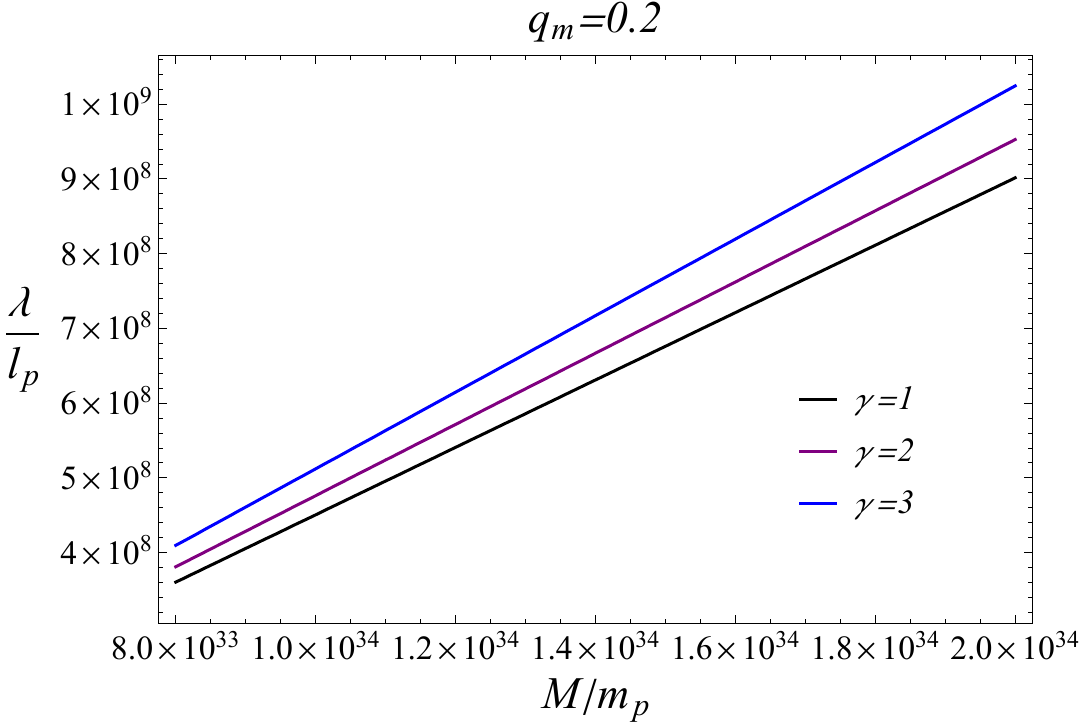}
\includegraphics[width=0.45\textwidth]{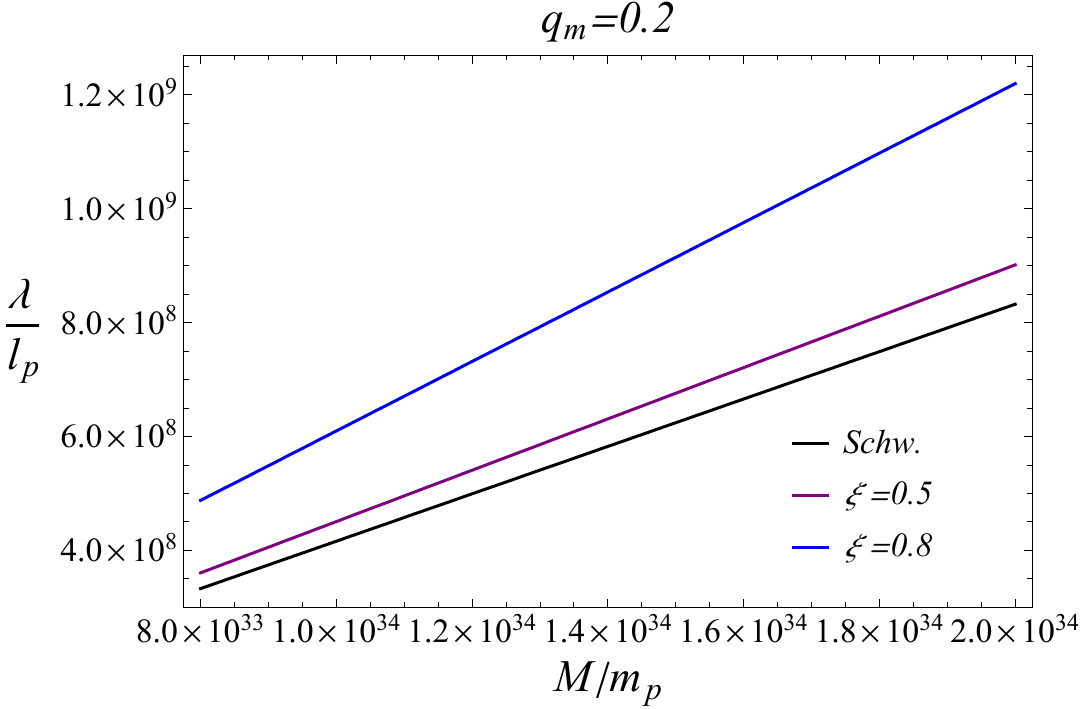}
\caption{\label{Fig.Wien}The peak Hawking wavelength (in units of Planck length) as a function of the black hole mass (in units of Planck mass) derived from Wien’s displacement law, comparing the Schwarzschild case and modified cases with parameters $\gamma$ and $\xi$.}
\end{figure*}

Fig. \ref{Fig.Wien} illustrates the variation of the peak radiation wavelength with black hole mass, as derived from Wien's displacement law. For all cases considered, the peak wavelength increases approximately linearly with mass, which is consistent with the standard expectation for black hole radiation: more massive black holes are colder and therefore emit radiation at longer wavelengths. In addition, as the parameters $\gamma$ and $\xi$ grow, the whole set of curves systematically shifts upward. This means that, for a fixed mass, the modified black holes have a longer peak wavelength, indicating a redshift of the spectrum towards the red. Such behavior originates directly from the reduction of the effective horizon temperature in modified gravity theories: a larger correction parameter leads to a stronger suppression of the temperature, which in turn causes the characteristic wavelength to be increasingly redshifted. These trends are also in agreement with the earlier discussion of excitation probabilities: a colder horizon implies a weaker population of high-frequency modes, thereby reducing the efficiency of horizon-driven radiative processes such as HBAR.

\section{Strong Gravitational Lensing by Degenerate Photon Sphere in the Quantum-Corrected Spacetime}\label{Sec.VI}

In static and spherically symmetric spacetimes, a degenerate photon sphere forms when an unstable photon sphere merges with a stable one. At that point, the second derivative of the effective potential vanishes, and the divergence behavior of the light deflection angle changes from logarithmic to power‑law. Such critical configurations typically occur in over‑charged or over‑extremal compact objects and often lack an event horizon. In this section, we will determine the critical parameter surface for the existence of a degenerate photon sphere in a quantum‑corrected metric, compute the divergence coefficient of the deflection angle in the strong‑deflection limit, and characterize it using local geometric quantities.

To describe the motion of massless test particles, we consider null geodesics. Exploiting spherical symmetry, we restrict attention to the equatorial plane $\theta=\pi/2$ without loss of generality. The geodesic equations are derived from the Lagrangian (\cite{UktamjonUktamov:2025qts,UktamjonUktamov:2025cso}):
\begin{eqnarray}\label{eq. Lagrangian}
\mathcal{L}=\frac{1}{2}\left[-f(r)\dot{t}^2+f(r)^{-1}\dot{r}^2+r^2\dot{\phi}^2\right]\,,
\end{eqnarray}
where dots denote derivatives with respect to an affine parameter. Since the metric does not depend on $t$ and $\phi$, we obtain two constants of motion: the energy $E=f(r)\dot{t}$ and the angular momentum $L=r^2\dot{\phi}$. For photons, the worldline satisfies the null condition $\mathcal{L}=0$. Substituting the constants yields the radial equation:
\begin{eqnarray}\label{eq. radial}
\dot{r}^2+f(r)\left[\frac{L^2}{r^2}-\frac{E^2}{f(r)}\right]=0\,,
\end{eqnarray}
to study the orbit shape, we introduce the impact parameter $b=\frac{L}{E}$ and change the derivative from the affine parameter to the azimuthal angle $\phi$. Using $\dot{\phi}=\frac{L}{r^2}$, we obtain:
\begin{eqnarray}\label{eq. radial phi}
\left(\frac{dr}{d\phi}\right)^2+V(r)=0\,,
\end{eqnarray}
where the effective potential $V(r)$ is defined as
\begin{eqnarray}\label{eq. potential}
V(r)=r^2f(r)\left[1-\frac{r^2}{b^2f(r)}\right]\,.
\end{eqnarray}

Physically, photons can only propagate in regions where $V(r)\leq0$. We focus on scattering trajectories, in which a photon comes in from infinity, approaches the central object until it reaches a minimum areal radius, and then recedes back to infinity. Such a trajectory possesses a single turning point at the radius of closest approach $r_0$, where the radial velocity vanishes. Accordingly, the turning point satisfies the condition that the effective potential is zero, i.e., $V(r_0)=0$. Using Eq.(\ref{eq. potential}), we obtain a direct relation between the impact parameter and the metric functions at the turning point:
\begin{eqnarray}\label{eq. impact}
b^2=\frac{r_0^2}{f(r_0)}\,.
\end{eqnarray}
Thus,we have obtained a direct relation between the impact parameter and the metric functions at the turning point. Substituting Eq.(\ref{eq. impact}) into Eq.(\ref{eq. potential}), we can write the potential as
\begin{eqnarray}\label{eq. potential2}
V(r)=r^2f(r)\left[1-\frac{r^2}{r_0^2}\frac{f(r_0)}{f(r)}\right]\,.
\end{eqnarray}

The total change in the azimuthal angle for such a scattering trajectory is obtained by integrating Eq.(\ref{eq. radial phi}) from $r_0$ to infinity. Because the trajectory is symmetric about its turning point, the full angular variation can be expressed as 
\begin{eqnarray}\label{eq. angle}
\Delta\varphi(r_0)=2\int_{r_0}^\infty\frac{dr}{\sqrt{-V(r)}}\,,
\end{eqnarray}
the deflection angle $\hat{\alpha}$ is defined as the deviation of the actual light ray from the straight-line path in flat spacetime, where the total change in azimuthal angle would be $\pi$; namely,
\begin{eqnarray}\label{eq. angle}
\hat{\alpha}=\Delta\varphi(r_0)-\pi\,,
\end{eqnarray}
the scattering trajectories are fully determined by the radius of turning point $r_0$ (or equivalently by the impact parameter $b$).

\section{Marginally unstable circular photon orbits}\label{Sec.VII}

This chapter aims to establish a strong‑deflection expansion theory for marginally unstable (degenerate) photon spheres. By combining the local analysis of the effective potential with the global integration of the deflection angle, we first present the mathematical conditions for degeneracy and analyse the asymptotic behaviour of the critical trajectory. Then, through the introduction of suitable small parameters, we separate the deflection‑angle integral into a divergent power‑law part and a finite regular part, thereby obtaining an analytic expansion for the deflection angle.

Circular orbits require $V(r_c)=0$ and $V^{\prime}(r_c)=0$, leading respectively to
\begin{eqnarray}\label{eq. circular1}
b_c^2=\frac{r_c^2}{f(r_c)}\,,\quad\quad\frac{f^{\prime}(r_c)}{f(r_c)}=\frac{2}{r}\,.
\end{eqnarray}
Hereafter, the subscript $c$ denotes evaluation at $r=r_c$. Stability is determined by the sign of $V^{\prime\prime}(r_c)$: $V^{\prime\prime}<0$ for unstable (ordinary photon sphere) and $V^{\prime\prime}>0$ for stable (anti‑photon sphere). Degeneracy occurs when these two types coalesce as a parameter is varied.

A degenerate photon sphere satisfies additionally $V^{\prime\prime}=0$. Using the explicit form of the effective potential, this condition becomes
\begin{eqnarray}\label{eq. circular2}
\frac{1}{r_c^2}-\frac{f^{\prime\prime}(r_c)}{2f(r_c)}=0\,.
\end{eqnarray}

\begin{figure}[ht!]
\includegraphics[width=0.45\textwidth]{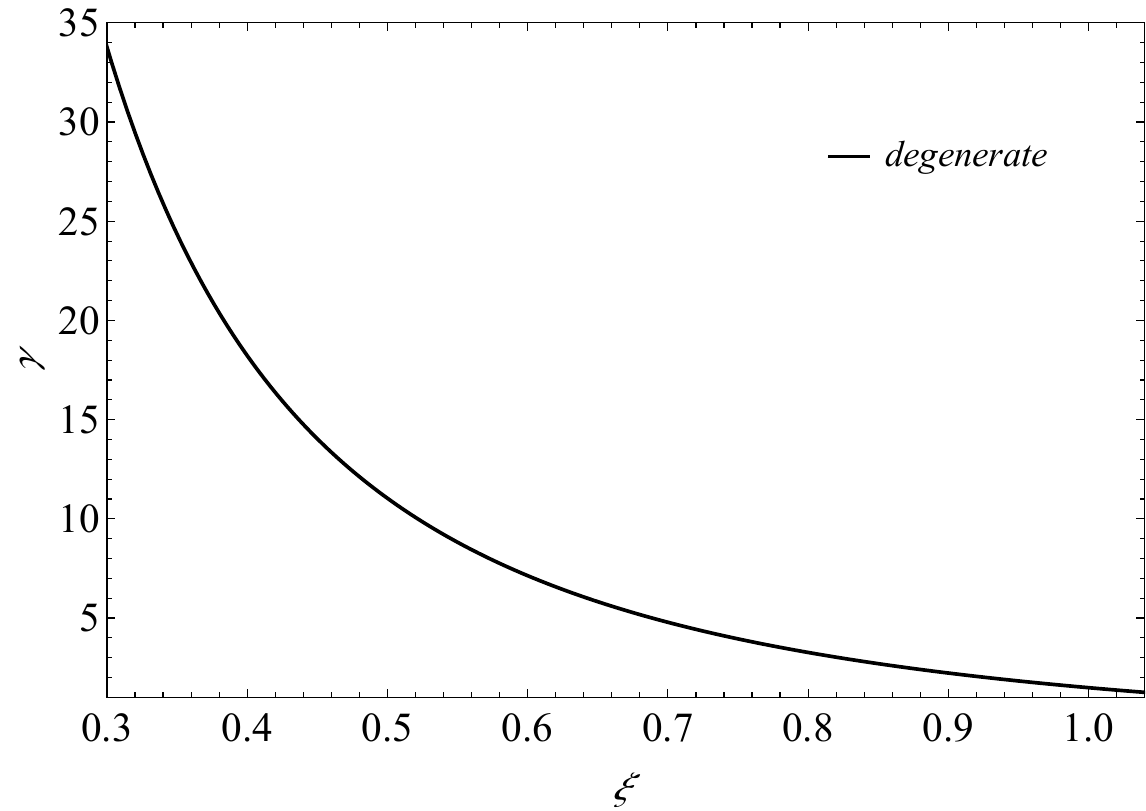}
\caption{The variation relationship between two parameters $\gamma$ and $\xi$ under the condition of satisfying the degenerate photon sphere.\label{Fig.degenerate}}
\end{figure}

Fig.\ref{Fig.degenerate} presents the critical parameter curve for which the degenerate photon sphere condition is satisfied. On this curve, the effective potential satisfies $V(r)=V^\prime(r)=V^{\prime\prime}(r)=0$, meaning that the unstable and stable photon spheres coalesce into a degenerate one. Only parameter combinations lying on this curve yield a marginally unstable photon sphere; away from the curve, the two photon spheres separate and the degeneracy condition no longer holds.

Near the degeneracy, the effective potential expands as
\begin{eqnarray}\label{eq. potental expand}
V(r)=\frac{1}{6}V_c^{\prime\prime\prime}(r-r_c)^3+\mathcal{O}(r^4)\,,
\end{eqnarray}
since the first and second derivatives vanish. For the critical trajectory with $b=b_c$, which asymptotically approaches $r_c$, we define the relative deviation $\delta_*=\frac{r}{r_c}-1$, Eq(\ref{eq. radial phi}) can be written as:
\begin{eqnarray}\label{eq. delta}
\left(\frac{d\delta_*}{d\phi}\right)^2=\kappa\delta_*^3\,,
\end{eqnarray}
where, $\kappa=-\frac{r_c}{6}V_c^{\prime\prime\prime}$. Because $V_c^{\prime\prime\prime}<0$, the solution is (\cite{Igata:2026hzb})
\begin{eqnarray}\label{eq. power-law}
\delta_*(\phi)\propto(\phi-\phi_0)^{-2}\,,
\end{eqnarray}
where $\phi_0$ is an integration constant. Since $\delta_*\to0$ only as $\lvert\phi-\phi_0\rvert\to\infty$, the critical trajectory winds around the marginally unstable circular orbit infinitely many times as it approaches $r_c$, indicating a power-law rather than exponential approach to the degenerate sphere.

\begin{figure*}[ht!]
\includegraphics[width=0.45\textwidth]{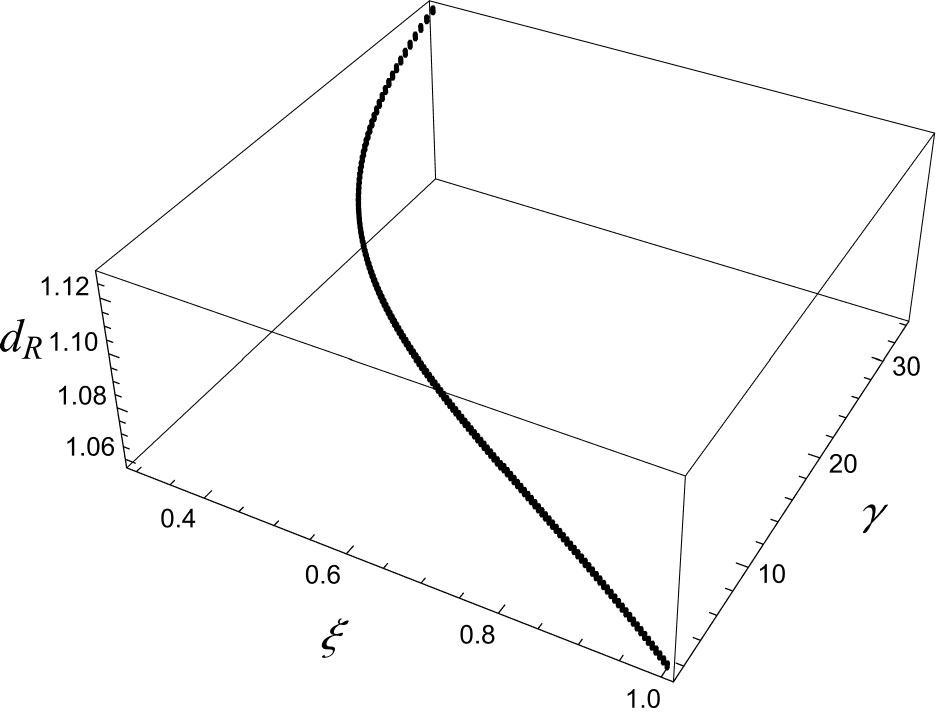}
\caption{Dependence of the constant term $d_R$ on the parameters $\gamma$ and $\xi$  extracted from high‑precision numerical integration. \label{Fig.dR}}
\end{figure*}

For a general scattering trajectory ($b\neq b_c$), the radius of closest approach $r_0$ satisfies
\begin{eqnarray}\label{eq. b}
b^2=\frac{r_0^2}{f(r_0)}\,,
\end{eqnarray}
Defining the small parameter $\delta=\frac{r_0}{r_c}-1$, and introduce the integration variable $z=1-\frac{r_0}{r}$ paves a way to rewrite  the change in azimuthal angle as:
\begin{eqnarray}\label{eq. anglez}
\Delta\phi(r_0)=2\int_0^1h(z;r_0)dz\,,
\end{eqnarray}
where
\begin{eqnarray}\label{eq. h}
h(z;r_0)=\left[(1-z)^4\frac{-V}{r_0^2}\right]^{-\frac{1}{2}}\,.
\end{eqnarray}
Near $z=0$, expanding the expression in the denominator yields the divergent part of the integral in the form
\begin{eqnarray}\label{eq. hexpand}
h(z;r_0)=\left(\zeta_1z+\zeta_2z^2+\zeta_3z^3\right)^{-\frac{1}{2}}\,,
\end{eqnarray}
where
\begin{eqnarray}\label{eq. zeta1}
\zeta_1&=&-\frac{V^{\prime}_0}{r_0}\,,\\\label{eq. zeta2}
\zeta_2&=&\frac{3V^{\prime}_0}{r_0}-\frac{V^{\prime\prime}_0}{2}\,,\\\label{eq. zeta3}\zeta_3&=&-\frac{3}{r_0}V^{\prime}_0+V^{\prime\prime}_0-\frac{r_0}{6}V^{\prime\prime\prime}_0\,.
\end{eqnarray}
In the degenerate limit $r_0\to r_c$(i.e.,$\delta\to0$), substituting Eqs. (\ref{eq. circular1}, \ref{eq. circular2}) into Eqs. (\ref{eq. zeta1}-\ref{eq. zeta3}), the above coefficients reduce to
\begin{eqnarray}\label{eq. zeta c1}
\zeta_1&=&3\kappa\delta^2\,,\\\label{eq. zeta c2}
\zeta_2&=&3\kappa\delta\,,\\\label{eq. zeta c3}
\zeta_3&=&\kappa\,.
\end{eqnarray}
Substituting the expansions of the coefficients $\zeta_i$ into the integral expression yields the leading term of azimuthal variation:
\begin{eqnarray}\label{eq. angle z}
I_D(r_0)=\frac{1}{\sqrt{\kappa}}\int_0^1\frac{2}{\sqrt{3\delta^2z+3\delta z^2+z^3}}dz\,.
\end{eqnarray}
To extract the power‑law divergence, we perform the scaling $z=\lvert\delta\rvert y$ and write $s=sgn(\delta)$. The leading term of the integral becomes the following:
\begin{eqnarray}\label{eq. angle y}
I_D(r_0)=\frac{1}{\sqrt{\kappa\lvert\delta\rvert}}\int_0^{1/\lvert\delta\rvert}F_s(y)dy\,,
\end{eqnarray}
with
\begin{eqnarray}\label{eq. Fsy}
F_s(y)=\frac{2}{\sqrt{y(y^2+3sy+3)}}\,.
\end{eqnarray}
In order to extract the leading divergent behavior and the accompanying constant term in the SDL, we decompose the integral into two contributions.
\begin{eqnarray}\label{eq. extract}
I_D(r_0)=\frac{1}{\sqrt{\kappa\lvert\delta\rvert}}\left(\int_0^{\infty}F_s(y)dy-\int_{1/\lvert\delta\rvert}^\infty F_s(y)dy\right)\,,
\end{eqnarray}
the first extends the upper limit to infinity, yielding a convergent constant, while the second is the tail integral, which can be evaluated asymptotically for large $y$. This decomposition is justified because, in the limit $\lvert\delta\rvert\to0$, the tail integral contributes a finite constant and does not affect the leading power‑law divergence.
Define
\begin{eqnarray}\label{eq. define}
\mathcal{U}_s&=&\int_0^\infty F_s(y)dy\,,\\
d_D&=&-\lim_{\lvert\delta\rvert\to0}\frac{1}{\sqrt{\kappa\lvert\delta\rvert}}\int_{1/\lvert\delta\rvert}^\infty F_s(y)dy\,.
\end{eqnarray}
The constant $\mathcal{U}_s$ depends only on the branch label $s$ and can be evaluated analytically in terms of gamma functions, while $d_D$ arises from the limiting contribution of the tail integral, which can be obtained from the large $y$ asymptotic expansion. Substituting these definitions into the previous decomposition, the divergent part of the integral takes the final form
\begin{eqnarray}\label{eq. IDr}
I_D(r_0)=\frac{\mathcal{U}_s}{\sqrt{\kappa\lvert\delta\rvert}}+d_D\,.
\end{eqnarray}
where
\begin{eqnarray}\label{eq. Us}
\mathcal{U}_+=\mathcal{U}\,,\quad\mathcal{U}_-=\sqrt{3}\mathcal{U}\,,
\end{eqnarray}
and
\begin{eqnarray}\label{eq. U}
\mathcal{U}=\frac{2\sqrt{\pi}\Gamma(1/6)}{3\Gamma(2/3)}\,.
\end{eqnarray}
Here, $\Gamma$ denotes the gamma function. Numerically, $\mathcal{U}\approx4.85730$ ($\mathcal{U}_-=\sqrt{3}\mathcal{U}\approx8.41309$). The finite constant $d_D$ arises from the tail of the integral in the large y region. Using the asymptotic behavior $F_s(y)\simeq2y^{-3/2}$ as $y\to\infty$, we obtain
\begin{eqnarray}\label{eq. dD}
d_D=-\frac{4}{\sqrt{\kappa}}\,.
\end{eqnarray}
It is worth noting that, since the large $y$ asymptotics of $F_s(y)$ are identical for $s=\pm1$, the constant $d_D$ is independent of the branch label s.
Having isolated the divergent contribution $I_D(r_0)$, we define the remaining part of the integral as the regular contribution
\begin{eqnarray}\label{eq. IR}
I_R(r_0)=2\int_0^1\left[h(z;r_0)-\frac{1}{\sqrt{\zeta_1z+\zeta_2z^2+\zeta_3z^3}}\right]dz\,,
\end{eqnarray}
where $h(z;r_0)$ is the exact integrand of the original integral. Since the two integrands share the same divergent behavior at $z=0$, their difference cancels the singularity, so $I_R(r_0)$ approaches a finite constant $d_R$ as $r_0\to r_c$, defined by
\begin{eqnarray}\label{eq. dR}
d_R\equiv\lim_{r_0\to r_c}I_R(r_0)\,
\end{eqnarray}
This (\ref{eq. dR}) constant cannot be obtained directly from the analytic expansion alone, but it can be determined via numerical integration, with results listed in the Fig. (\ref{Fig.dR}).
Combining the above results, the azimuthal change can be written as
\begin{eqnarray}\label{eq. com angle}
\hat{\alpha}(r_0)=c_s\lvert\delta\rvert^{-1/2}+d+\mathcal{O}(\lvert\delta\rvert^{1/2})\,,
\end{eqnarray}
where
\begin{eqnarray}\label{eq. cs}
c_s&=&\frac{\mathcal{U}_s}{\sqrt{\kappa}}\,,\\\label{eq. d}
d&=&d_D+d_R-\pi\,.
\end{eqnarray}
If we instead use the impact‑parameter deviation $\epsilon=\frac{b}{b_c}-1$, and employ the near‑critical relation
\begin{eqnarray}\label{eq. cs}
\epsilon=\mathcal{K}\delta^3+\mathcal{O}(\delta^4)\,.
\end{eqnarray}
with 
\begin{eqnarray}\label{eq. K}
\mathcal{K}\equiv-\frac{r_cV^{\prime\prime\prime}_c}{12f(r_c)}\,,
\end{eqnarray}
which is a positive local geometric quantity that measures the third‑order conversion coefficient between the impact‑parameter deviation and the closest‑approach deviation. Substituting this relation into the expansion for $\hat{\alpha}(r_0)$, we rewrite the deflection angle in terms of the impact‑parameter deviation as
\begin{eqnarray}\label{eq. angle a}
\hat{\alpha}(b)=\frac{\bar{c_s}}{\rvert\epsilon\lvert^{1/6}}+d+\mathcal{O}(\rvert\epsilon\lvert^{1/6})\,.
\end{eqnarray}
where
\begin{eqnarray}\label{eq. barcs}
\bar{c_s}=\mathcal{U}_s\left[2\kappa^2\left(1-\frac{2m_c}{r_c}\right)\right]^{-1/6}\,.
\end{eqnarray}
Here, $m_c\equiv m(r_c)$ is the Misner–Sharp mass evaluated in the critical radius, related to the metric functions through $m_c=\frac{1-f(r_c)}{2}r_c$, and characterizes the local compactness at the critical point. This shows that the deflection angle diverges as a $-1/6$ power law in the impact ‐ parameter space, equivalent to the $-1/2$ power law in the closest ‐ approach space.

Fig.\ref{Fig.dR} shows the dependence of the finite constant term $d_R$ on the parameters $\gamma$ and $\xi$. The overall trend indicates that $d_R$ increases monotonically with $\gamma$, while it decreases monotonically as $\xi$ grows. This opposite dependence reflects the distinct roles of the two modification parameters in the metric near the degenerate photon sphere: An increase in $\gamma$ enhances the contribution of higher‑order curvature corrections to the photon binding effect, thereby increasing the value of $d_R$; as $\xi$ increases, the quantum gravity corrections gradually become stronger, and these corrections tend to reduce the local curvature effects near the degenerate photon sphere, leading to a monotonic decrease in $d_R$. The opposite algebraic signs of their contributions to the third derivative of the effective potential give rise to the observed trend of $d_R$ increasing with $\gamma$ and decreasing with $\xi$.

From Tables \ref{table.gamma} and \ref{table.xi}, it is evident that the physical quantities associated with the degenerate photon sphere exhibit systematic monotonic variations with the two modification parameters $\gamma$ and $\xi$. As $\xi$ increases (Table \ref{table.gamma}), the corresponding $\gamma$ decreases; the critical radius $r_c$, critical impact parameter $b_c$, the absolute value of the third derivative $\lvert V^{\prime\prime\prime}_c\rvert$, $\kappa$ and $\mathcal{K}$ all decrease, while the divergence coefficients $c_+$ and $\bar{c}_+$ increase. Conversely, as $\gamma$ increases (Table \ref{table.xi}), the corresponding $\xi$ decreases; $r_c$, $b_c$, $\lvert V^{\prime\prime\prime}_c\rvert$, $\kappa$ and $\mathcal{K}$ all increase, whereas $c_+$ and $\bar{c}_+$ decrease. Furthermore, the tail constant $d_D$ exhibits a negative correlation with $\xi$ and a positive correlation with $\gamma$, while the regular‑part limit $d_R$ shows the opposite behavior. Overall, $\gamma$ and $\xi$ play opposing regulatory roles in determining the local geometry and divergence coefficients near the degenerate photon sphere.

\begin{table*}[htbp]
    \centering
    \caption{The values of degenerate photon sphere parameters for varying $\xi$ and fixed $\gamma$ (determined by the degeneracy condition).\label{table.gamma}}
    \begin{tabular*}{\textwidth}{@{\extracolsep{\fill}} l@{\hspace{1.5em}}|c c c c c c c c c c c}
    \toprule
    $\xi$ & $\gamma$ & $r_c$ & $b_c$ & $m_c$ & $V^{\prime\prime\prime}_c$ & $\kappa$ & $\mathcal{K}$ & $c_+$ & $\bar{c}_+$ & $d_D$ & $d_R$\\[4pt]
    \hline
    0.4
         & $18.23$
         & $2.15$
         & $22.02$
         & $0.85$
         & $-2.85$
         & $1.02$
         & $2.44$
         & $4.81$
         & $5.58$
         & $-3.96$
         & $1.12$\\[4pt]
    0.6
         & $7.13$
         & $2.12$
         & $21.69$
         & $0.84$
         & $-2.80$
         & $0.99$
         & $2.39$
         & $4.89$
         & $5.65$
         & $-4.02$
         & $1.11$\\[4pt]
    0.8
         & $3.26$
         & $2.08$
         & $21.21$
         & $0.83$
         & $-2.72$
         & $0.94$
         & $2.31$
         & $5.01$
         & $5.76$
         & $-4.13$
         & $1.09$\\[4pt]
    1.0
         & $1.49$
         & $2.02$
         & $20.56$
         & $0.81$
         & $-2.59$
         & $0.87$
         & $2.20$
         & $5.21$
         & $5.94$
         & $-4.29$
         & $1.05$\\
    \hline
    \end{tabular*}
    \label{tab:placeholder}
\end{table*}

\begin{table*}[htbp]
    \centering
    \caption{The values of degenerate photon sphere parameters for varying $\gamma$ and fixed $\xi$(determined by the degeneracy condition).\label{table.xi}}
    \begin{tabular*}{\textwidth}{@{\extracolsep{\fill}} l@{\hspace{1.5em}}|c c c c c c c c c c c}
    \toprule
    $\gamma$ & $\xi$ & $r_c$ & $b_c$ & $m_c$ & $V^{\prime\prime\prime}_c$ & $\kappa$ & $\mathcal{K}$ & $c_+$ & $\bar{c}_+$ & $d_D$ & $d_R$\\[4pt]
    \hline
    1
         & $1.09$
         & $1.98$
         & $20.20$
         & $0.80$
         & $-2.51$
         & $0.83$
         & $2.14$
         & $5.33$
         & $6.05$
         & $-4.39$
         & $1.04$\\[4pt]
    5
         & $0.69$
         & $2.10$
         & $21.49$
         & $0.83$
         & $-2.77$
         & $0.97$
         & $2.36$
         & $4.93$
         & $5.69$
         & $-4.06$
         & $1.11$\\[4pt]
    10
         & $0.52$
         & $2.13$
         & $21.83$
         & $0.84$
         & $-2.82$
         & $1.00$
         & $2.41$
         & $4.85$
         & $5.62$
         & $-4.00$
         & $1.12$\\[4pt]
    15
         & $0.44$
         & $2.14$
         & $21.97$
         & $0.85$
         & $-2.85$
         & $1.02$
         & $2.43$
         & $4.82$
         & $5.59$
         & $-3.97$
         & $1.13$\\
    \hline
    \end{tabular*}
    \label{tab:placeholder}
\end{table*}

\section{Conclusions}\label{Sec.VIII}

In this work, we investigate two complementary probes of strong gravity in the context of quantum-corrected black holes: (i) an operational near-horizon quantum channel, specifically the HBAR mechanism, generated by infalling two-level atoms interacting with field modes near the horizon, and (ii) a strong-deflection expansion for null-geodesic scattering in the vicinity of a degenerate photon sphere.

We have examined a static, spherically symmetric quantum-corrected black hole, mapping the horizon-structure dependence on the correction parameters, identifying the extremal boundary between black-hole and no-horizon regimes, and analyzing the Hawking temperature, which vanishes at the extremal limit.

Subsequently, by identifying the near-critical region controlling the divergence, rescaling the integration variable, and nonsingularity extracting the near-orbit contribution to the deflection-angle integral at marginality, we have obtained the leading power-law coefficient and separated the finite offset into a universal near-region term and a quantum correction-dependent regular part.

On the quantum side, we have developed the near-horizon formalism required for the HBAR implementation. By reducing the field dynamics to the near-horizon sector, we have shown that infalling atomic detectors respond according to a conformal near-horizon structure, yielding a thermal spectrum at the horizon temperature. We then used a Lindblad master equation for the radiation field, showed that a thermal steady state arises, and derived a consistent entropy–energy flux relation for HBAR that parallels the thermodynamics of horizon radiation. We have derived a Wien-type displacement law for the HBAR spectrum, directly connecting the peak wavelength to horizon thermodynamic data, thereby offering a further observable signature of quantum-corrected black holes via near-horizon radiation.

Several directions for generalization follow naturally from this study. On the HBAR side, possible extensions include going beyond the dominant sector and dilute limit, incorporating additional mode structure, and including greybody factors and backreaction for quantitative flux predictions. Regarding the strong-deflection expansion, we leave for future study a thorough exploration of higher-order degeneracies and their associated power-law scaling exponents.
Extending the present analysis to rotating regular black holes would be particularly valuable, as it would enable a simultaneous comparison of spin-dependent photon-sphere observables with the operational HBAR channel. Overall, we have shown that quantum-corrected black holes support a joint description in which strong-deflection expansion of the deflection angle near a degenerate photon sphere analysis and HBAR thermodynamics are both controlled by the same horizon-scale data, thus providing a concrete avenue for testing nonsingular quantum-correction models with present and future horizon-scale observations.

\begin{acknowledgments}
This research was funded by the National Natural Science Foundation of China (NSFC) under Grant No. U2541210. 
\end{acknowledgments}

\appendix

\section{Useful expressions}\label{sec.app}

We present useful expressions used in the paper below.

The variables introduced for Eqs.(\ref{eq. derivative1},\ref{eq. derivative2}) are :
\begin{widetext}
\begin{subequations}
\begin{align}
    &\mathcal{A}_1=\xi^2+\frac{12r^5+(r+M\gamma)\xi^4}{\sqrt{4r^6+(r+M\gamma)^2\xi^4}}\,,\\
    &\mathcal{B}_1=(r+M\gamma)\xi^2+\sqrt{4r^6+(r+M\gamma)^2\xi^4}\,,\\
    &\mathcal{C}_1=1/\sqrt{4r^6+(r+M\gamma)^2\xi^4}\,.
\end{align}
\end{subequations}
\end{widetext}
The variables introduced for Eqs.(\ref{eq. derivatives rh},\ref{eq. derivatives rh2}) are :

\begin{widetext}
\begin{subequations}
    \begin{align}            &\mathcal{A}_2=\xi^2+\frac{12r^5+\xi^4\left(r+\frac{\gamma\mathcal{C}_2}{4r^2-2\gamma\xi^2}\right)}{\sqrt{4r^6+\xi^4\left(r+\frac{\gamma\mathcal{C}_2}{4r^2-2\gamma\xi^2}\right)^2}}\,,\\
            &\mathcal{B}_2=\xi^2\left(r+\frac{\gamma\mathcal{C}_2}{4r^2-2\gamma\xi^2}\right)\,,\\
            &\mathcal{C}_2=r\xi^2+\sqrt{r^2(4r^4-2r^2\gamma\xi^2+\xi^4)}\,,\\
          &\mathcal{D}_2=-\frac{\left[48r^7-24r^5\gamma\xi^2+4r^3\xi^4-r\gamma\xi^6+\gamma\xi^4\sqrt{r^2(4r^4-2r^2\gamma\xi^2+\xi^4)}\right]^2}{(4r^2-2\gamma\xi^2)^2\left[4r^6+\xi^4\left(r+\frac{\gamma\mathcal{C}_2}{4r^2-2\gamma\xi^2}\right)^2\right]^{3/2}}\\
            &+\frac{60r^4+\xi^4}{\sqrt{4r^6+\xi^4\left(r+\frac{\gamma\mathcal{C}_2}{4r^2-2\gamma\xi^2}\right)^2}}\,.
    \end{align}
    \end{subequations}
\end{widetext}

\allowdisplaybreaks


\bibliography{mybib}    
\end{document}